\documentclass[manuscript]{aastex}

\newcommand{\myemail}{tanaka@vega.ess.sci.osaka-u.ac.jp}

\shorttitle{SPECTRAL EVOLUTION OF YOUNG TEV PWNE}
\shortauthors{S. J. TANAKA /& F. TAKAHARA}

\begin{document}

\title{STUDY OF FOUR YOUNG TEV PULSAR WIND NEBULAE WITH A SPECTRAL EVOLUTION MODEL}

\author{Shuta J. Tanaka\altaffilmark{1} and Fumio Takahara}
\affil{Department of Earth and Space Science, Graduate School of Science, Osaka University, 1-1 Machikaneyama-cho, Toyonaka, Osaka 560-0043, Japan }

\altaffiltext{1}{e-mail: \myemail}

\begin{abstract}
We study four young Pulsar Wind Nebulae (PWNe) detected in TeV $\gamma$-rays, G21.5-0.9, G54.1+0.3, Kes 75, and G0.9+0.1, using the spectral evolution model developed and applied to the Crab Nebula in our previous work.
We model the evolution of magnetic field and particle distribution function inside a uniformly expanding PWN considering a time-dependent injection from the pulsar and radiative and adiabatic losses.
Considering uncertainties in the interstellar radiation field (ISRF) and their distance, we study two cases for each PWN.
Because TeV PWNe have a large TeV $\gamma$-rays to X-rays flux ratio, the magnetic energy of the PWNe accounts for only a small fraction of the total energy injected (typically a few $\times 10^{-3}$).
The $\gamma$-ray emission is dominated by inverse Compton scattering off the infrared photons of the ISRF.
A broken power-law distribution function for the injected particles reproduces the observed spectrum well, except for G0.9+0.1.
For G0.9+0.1, we do not need a low energy counterpart because adiabatic losses alone are enough to reproduce the radio observations.
High energy power-law indices at injection are similar (2.5 -- 2.6), while low energy power-law indices range from 1.0 to 1.6.
The lower limit of the particle injection rate indicates that the pair multiplicity is larger than $10^4$.
The corresponding upper limit of the bulk Lorentz factor of the pulsar winds is close to the break energy of the broken power-law injection, except for Kes 75.
The initial rotational energy and the magnetic energy of the pulsars seem anticorrelated, although the statistics are poor.
\end{abstract}


\section{INTRODUCTION}\label{intro}
A pulsar wind nebula (PWN) is created by the interaction between the pulsar wind and the surrounding supernova (SN) ejecta \citep[][]{rg74, kc84}.
The PWN is a shocked pulsar wind composed of a relativistic non-thermal electron-positron plasma and magnetic fields.
As a result, a PWN shines from radio through TeV $\gamma$-rays via synchrotron radiation and inverse Compton scattering.
The observed spectrum of the PWN gives us important information on the pair production multiplicity $\kappa$ of the central pulsar \citep[][]{d07}, the magnetization parameter $\sigma$ in the pulsar wind immediately upstream the termination shock \citep[][]{kc84}, and the particle acceleration process at the pulsar wind termination shock \citep[e.g., ][]{fz10}.
It should be noted that the radiation spectrum depends on time through the evolution of the energy injection, the expansion of the PWN, and radiative and adiabatic losses \citep[e.g., ][]{tt10, get09, bet10c}.

Through the recent development of $\gamma$-ray observations, many PWNe have been discovered in $\gamma$-rays and many detailed structures of them have also been found in other wavelengths \citep[e.g., HESS J1640-465 studied by][]{aet06, let09}.
Although the observed PWNe show common characteristics, such as flat radio spectrum and X-ray spectral steepening with distance from the pulsar \citep[c.f., ][]{gs06}, the differ.
We discuss these individual differences of PWNe in this paper.
They include the age and the spin-down power of the central pulsar, the observed size of the PWN and the TeV $\gamma$-rays to X-rays flux ratio.
Here we study these characteristics because they constrain the physical condition of the pulsar magnetosphere and the pulsar wind.

We built a spectral evolution model of PWNe in our previous work \citep[][]{tt10}.
The application of the model to the Crab Nebula well reproduces the current observed spectrum and the radio flux evolution.
In this paper, we apply this model to four young PWNe detected in TeV $\gamma$-rays, G21.5-0.9, G54.1+0.3, Kes 75, and G0.9+0.1.
We pick up these four PWNe according to three criteria.
(1) They reveal observed non-thermal spectrum at least in radio, X-rays, and TeV $\gamma$-rays.
(2) They have a central pulsar with known period and its derivative.
(3) They have an almost spherical shape with a known angular extent, i.e., they are young enough not to reveal signatures of interaction with SNR reverse shock.
For the application to the PWNe other than the Crab Nebula, we include the interstellar radiation field (ISRF) in infrared and optical bands as the target photons of the inverse Compton scattering.

We assume a constant expansion velocity of the PWN.
This assumption is appropriate to young PWNe.
\citet{get09} considered the dynamical evolution of a PWN inside a supernova remnant (SNR).
Their model of the PWN expansion gives a more realistic dynamical evolution than the constant expansion velocity and also applies to old PWNe.
However, their expansion model includes many unknown parameters to determine the dynamical evolution including the energy of the SN explosion, the mass of the SN ejecta and the density of the surrounding interstellar medium.
We factor these uncertain quantities into one parameter: the constant expansion velocity.
Note that a simple estimate of the radius of the PWN in the early phase of its evolution, $R_{\rm PWN} \propto t^{6/5}$ \citep[e.g., ][]{vet01}, is close to an expansion at constant velocity.

In Section \ref{model}, we describe our model and slightly improve it for the application to young TeV PWNe other than the Crab Nebula.
In Sections \ref{g21.5} -- \ref{g0.9}, we apply the model to G21.5-0.9, G54.1+0.3, Kes 75, and G0.9+0.1 and discuss about their individual characteristics.
A comparative discussions about the young TeV PWNe we studied, including the Crab Nebula and the conclusions are made in Section \ref{dis_con}.

\section{THE MODEL}\label{model}
We give here a brief description of our spectral evolution model of PWNe \citep[][]{tt10}.
For application to young TeV PWNe other than the Crab Nebula, we improve the target photon fields for inverse Compton scattering.
We include the ISRF in infrared and optical bands, which were ignored in the previous work.
Lastly, we describe the fitting procedure.

\subsection{Basic Ingredients of Our Model}
We use a one-zone model of PWNe.
A PWN is a uniform sphere expanding at a constant velocity $v_{\rm{PWN}}$, i.e., the radius of PWN is given by $R_{\rm PWN}(t) = v_{\rm PWN} t$.
The contents of the PWN are the magnetic field and the relativistic non-thermal electron-positron plasma and they are injected from the central pulsar.
The magnetic energy injection $\dot{E}_{\rm mag}(t) = \eta L(t)$ and the particle energy injection $\dot{E}_{\rm part}(t) = (1 - \eta) L(t)$ are characterised by the fraction parameter $\eta$ ($0 \le \eta \le 1$) and the spin-down power $L(t)$.
The evolution of the spin-down power $L(t)$ is given by
\begin{equation}\label{eq_spin-down}
L(t) = L_{\rm 0} \left( 1 + \frac{t}{\tau_{0}} \right)^{-\frac{n+1}{n-1}},
\end{equation}
where $L_0$ is the initial spin-down power and $\tau_0$ is the spin-down time.
For $t > \tau_0$, the total energy injected into the PWN up to time $t$ ($E_{\rm tot}(t) = \int^{t}_0 L(t') dt'$) approximately corresponds to the initial rotational energy of the central pulsar $L_0 \cdot \tau_0 = I \Omega^2_0 / (n - 1)$, where $\Omega_0 = 2 \pi / P_0$ is the initial angular velocity.
For $t < \tau_0$, $E_{\rm tot}$ is smaller than $L_0 \cdot \tau_0$. 
We need four quantities, the current pulsar period $P$, its time derivative $\dot{P}$, braking index $n$ and the age of pulsar $t_{\rm{age}}$, to fix the evolution of the spin-down power, assuming that the moment of inertia of the pulsar is $10^{45}\rm g \cdot \rm cm^2$.
Note that three time scales appear in the spin-down evolution of the pulsar: the age of pulsar $t_{\rm age}$, the spin-down time $\tau_0$ and the characteristic age $\tau_{\rm c}$.
They are related to each other through the simple relation
\begin{equation}\label{eq_age_tc_t0}
\tau_{\rm c} = \frac{n - 1}{2} (\tau_0 + t_{\rm age}).
\end{equation}
The customarily used braking index $n = 3$ gives $\tau_{\rm c} = \tau_0 + t_{\rm age}$.

We assume that the distribution of the particles in the PWN is isotropic, and then the evolution of the particle distribution $N(\gamma, t)$ is given by the continuity equation in energy space,
\begin{equation}\label{eq_continuity}
\frac{ \partial}{ \partial t} N(\gamma, t) + \frac{ \partial}{ \partial \gamma} \left( \dot{\gamma}(\gamma, t) N(\gamma, t) \right) = Q_{\mathrm{inj}}(\gamma, t).
\end{equation}
We assume that the particle injection $Q_{\rm inj}(\gamma, t)$ follows the broken power-law distribution
\begin{equation}\label{eq_injection}
Q_{\rm inj}(\gamma, t) = \left\{
\begin{array}{ll}
Q_{\rm 0}(t) (\gamma / \gamma_{\rm {b}})^{-p_{\rm 1}} & \mbox{ for $\gamma_{\rm min} \leq \gamma \leq \gamma_{\rm b}  $ ,} \\
Q_{\rm 0}(t) (\gamma / \gamma_{\rm {b}})^{-p_{\rm 2}} & \mbox{ for $\gamma_{\rm b}   \leq \gamma \leq \gamma_{\rm max}$ ,}
\end{array} \right.
\end{equation}
where $\gamma$ is the Lorentz factor of the relativistic electrons and positrons.
We introduce the parameters of the injection spectrum, $\gamma_{\rm min}$, $\gamma_{\rm b}$, $\gamma_{\rm max}$, $p_{\rm 1} < 2$ and $p_{\rm 2} > 2$, which are the minimum, break and maximum Lorentz factors and the power-law indices at the low and high energy ranges of the injection spectra, respectively.
We require that the normalization $Q_{\rm{0}}(t)$ satisfies $(1-\eta)L(t) = \int_{\gamma_{\rm min}}^{\gamma_{\rm max}} Q_{\rm inj}(\gamma, t) \gamma m_{\rm e} c^{2} d\gamma$, where $m_{\rm e}$ and $c$ are the mass of an electron (or positron) and the speed of light, respectively.
We consider the cooling effects of the relativistic particles $\dot{\gamma}(\gamma, t)$ including the synchrotron radiation $\dot{\gamma}_{\mathrm{syn}}(\gamma,t)$, the inverse Compton scattering off the ISRF $\dot{\gamma}_{\mathrm{IC}}(\gamma)$ and the adiabatic expansion $\dot{\gamma}_{\mathrm{ad}}(\gamma,t)$.
The detailed description of the ISRF will be discussed in Section \ref{background}.
The cooling time of the particles $\tau_{\rm cool}(\gamma, t) = \gamma / \left| \dot{\gamma}(\gamma, t) \right|$ is an important time-scale in addition to $t_{\rm age}$ and $\tau_0$, and these three time-scales characterize the evolution of the particle distribution.

For the magnetic field evolution, we assume the following form of the magnetic field energy conservation,
\begin{equation}\label{eq_magnetic-field}
\frac{4\pi}{3} (R_{\rm{PWN}}(t))^{3} \cdot \frac{(B(t))^{2}}{8\pi} = \int_0^{t} \eta L(t') dt' =  \eta E_{\rm tot}(t).
\end{equation}
The magnetic field approximately evolves as $B(t) \propto t^{-1}$ for $t < \tau_0$ and $B(t) \propto t^{-1.5}$ for $t > \tau_0$.
Note that some justifications of this magnetic field evolution model are discussed in the previous paper \citep[see Section 2.2 of ][]{tt10}.

For the calculation of the radiation spectrum, we assume that the radiation is isotropic.
The radiation processes which we consider are synchrotron radiation and inverse Compton scattering off the synchrotron radiation (SSC) and the ISRF including the cosmic microwave background (IC/CMB) radiation, infrared photons from dust grains (IC/IR) and optical photons from starlight (IC/OPT).

Lastly, we mention the evolution of the synchrotron cooling break frequency for convenience in the later discussion.
Assuming that the inverse Compton cooling is ineffective in most of the evolutionary phase \citep[e.g., Figure 4 of ][]{tt10}, we divide the particle distribution into two populations; one corresponds to the high energy particles which lose energy by synchrotron cooling, the other is the low energy particles which lose energy by adiabatic cooling.
The critical energy $\gamma_{\rm c}(t)$ is defined by equating the synchrotron cooling time $\tau_{\rm syn}(\gamma, t)$ with the adiabatic cooling time $\tau_{\rm ad}(\gamma, t) = t$, which is:
\begin{equation}\label{eq_break-energy}
\gamma_{\rm c}(t) = \frac{6 \pi m_{\rm e} c}{\sigma^{}_{\rm T} t B^2(t)} \sim 2.45 \times 10^8 \left( \frac{t}{1 \rm kyr} \right)^{-1} \left( \frac{B(t)}{10 \mu \rm G} \right)^{-2}.
\end{equation}
For $\gamma > \gamma_{\rm c}(t)$, the synchrotron cooling dominates over the adiabatic cooling and vice versa for $\gamma < \gamma_{\rm c}(t)$. 
The characteristic frequency of synchrotron radiation $\nu_{\rm syn}$ is a function of the particle energy and the magnetic field.
Because of the rapid decrease of the magnetic field strength, the synchrotron cooling break frequency $\nu_{\rm c}(t) \equiv \nu_{\rm syn}(\gamma_{\rm c}(t), B(t))$ increases with time, which is: 
\begin{equation}\label{eq_break-frequency}
\nu_{\rm c}(t) \sim 1.22 \times 10^{17} {\rm Hz} \left( \frac{\gamma_{\rm c}(t)}{10^8} \right)^2 \left( \frac{ B(t)}{10 \mu \rm G} \right) \propto \left\{
\begin{array}{ll}
t & \mbox{ for $t < \tau_0$ ,} \\
t^{2.5} & \mbox{ for $t > \tau_0$ .} \\
\end{array} \right.
\end{equation}

\subsection{Galactic Interstellar Radiation Field as the Target of Inverse Compton Scattering}\label{background}
We use the ISRF model described by \citet{pet06} as a reference.
Their ISRF model depends on the distance from the galactic center $r$ and the height from the galactic plane $z$.
Although the contribution of localized dusts and stars to the ISRF changes the actual ISRF around the objects, we ignore the possibility of local effects, except for the model 2 of G54.1+0.3.

We assume that the ISRF has three components.
The CMB is a blackbody radiation with a temperature $T_{\rm CMB} = 2.7 \rm K$.
The spectra in infrared and optical bands are modified blackbodies characterized by a temperatures and an energy density $(T_{\rm IR}, U_{\rm IR})$ and $(T_{\rm OPT}, U_{\rm OPT})$, respectively.
Because the temperature dependence of the IC/IR ($T_{\rm IR} \sim 30 - 50K$) and the IC/OPT ($T_{\rm OPT} \sim 3500 - 4000K$) is weak, we fix $T_{\rm IR} = 40 \rm K$ (mildly Klein-Nishina regime), and $T_{\rm OPT} = 4000 \rm K$ (mostly Klein-Nishina regime) for all objects.
On the other hand, the IC/IR and the IC/OPT luminosities strongly depend on the energy densities $U_{\rm IR}$ and $U_{\rm OPT}$ which vary with galactic locations.
From the distance to each PWN and its galactic coordinates, we can roughly estimate $(r, z)$ for each object.
Adopted values for each object are listed in Table \ref{tbl-1}.

\subsection{Fitting Procedure}\label{procedure}
Here, we summarize the model parameters and describe the fitting procedure.
We adopt the distance to the object $d$, the angular extent of the object, and the central pulsar parameters $P$, $\dot{P}$ and $n$ from observations.
Once the distance to the object is fixed, we can determine the radius of the object $R_{\rm PWN}$ and the energy densities of the ISRF $U_{\rm IR}$ and $U_{\rm OPT}$. 

Fitting parameters are the age $t_{\rm age}$ (assuming that the age of the central pulsar is the same as that of the PWN), the fraction parameter $\eta$, and the parameters of the injection spectrum in Equation (\ref{eq_injection}), $\gamma_{\rm max}$, $\gamma_{\rm b}$, $\gamma_{\rm min}$, $p_1$ and $p_2$.
In our spectral evolution model, the age $t_{\rm age}$ and the fraction parameter $\eta$ of PWNe are primarily determined by the absolute power of the synchrotron radiation and the power ratio of the inverse Compton scattering to the synchrotron radiation \citep[see Section 3.3 of ][]{tt10}.
The parameters of the injection spectrum in Equation (\ref{eq_injection}) are determined by comparing the calculated spectrum with the detailed shape of the observed spectrum \citep[see Section 3.4 of ][]{tt10}.
The parameters $\gamma_{\rm min}$ and $\gamma_{\rm max}$ are relatively unconstrained.
We obtain only upper limit of $\gamma_{\rm min}$ from the lowest frequency in radio observations.
We choose lower limit of $\gamma_{\rm max}$ so that $\nu_{\rm syn} (\gamma_{\rm max}, B)$ becomes an order of magnitude larger than the observed highest frequency in X-rays, since no clear spectral rollover has been observed for the four PWNe.

The fitted parameters which reproduce the multi-wavelength observations are almost uniquely determined except for $\gamma_{\rm min}$ and $\gamma_{\rm max}$, if we fix the distance to the PWNe and the energy densities of the ISRF.
To understand effects of uncertainties in the observed flux, measured distance and the ISRF model on our calculations, we study two different models for each PWN in Sections \ref{g21.5} -- \ref{g0.9}.

The adopted and fitted parameters gives the values of the expansion velocity $v_{\rm PWN}$, the current magnetic field strength of the PWNe $B_{\rm now}$ and the parameters of the central pulsars $\tau_0$ and $L_0 \cdot \tau_0$.
The adopted, fitted and derived values for each model are listed in Table \ref{tbl-1}.

\section{G21.5-0.9}\label{g21.5}
G21.5-0.9 is a composite SNR and its PWN is observed in radio \citep[][]{set89, bet01}, infrared \citep[][]{gt98}, X-rays \citep[][]{tet10b, det09b}, and TeV $\gamma$-rays \citep[][]{det07b}.
In X-rays, G21.5-0.9 consists of an outer halo $\sim 150''$ in radius, which is the SNR component, and a PWN $\sim 40''$ in radius \citep[][]{ms05, bet05, ms10}.
In radio, the outer halo is not observed and the size of the PWN is comparable with that in the X-rays \citep[][]{bet01, bb08, bet10a}.
Infrared observations of PWNe are difficult and G21.5-0.9 is one of a few PWNe whose non-thermal infrared spectrum is observed \citep[][]{gt98}.
{\it Fermi} LAT puts an upper limit on the flux of the PWNe component in GeV $\gamma$-rays \citep[][]{aet11}.
The central pulsar of G21.5-0.9 (PSR J1833-1034) is observed in radio and GeV $\gamma$-rays with a period $P = 6.19 \times 10^{-2} \rm sec$, its time derivative $\dot{P} = 2.02 \times 10^{-13} \rm sec \cdot sec^{-1}$ and unknown braking index (we assume $n = 3$) \citep[][]{cet06, aet09a}.
The characteristic age $\tau_{\rm c}$ of the central pulsar is $4.9 \rm kyr$, but \citet{bb08} suggested $t_{\rm age} \sim 900 \rm yr$ from the observation of the expansion rate of the PWN.
On the other hand, \citet{wet86} suggested that G21.5-0.9 might be the historical supernova in 48 BC, i.e., $t_{\rm age} \sim 2 \rm kyr$.
We assume that the distance to G21.5-0.9 is 4.8kpc \citep[][]{tl08} and then we approximate G21.5-0.9 PWN as a sphere of radius 1.0pc at $(r, z) \sim (4 {\rm kpc}, 80 {\rm pc})$.
Lastly, we get $(U_{\rm IR}, U_{\rm OPT}) = (1.0 {\rm eV/cm^3}, 2.0 {\rm eV/cm^3})$ as the energy density of the ISRF.
Considering that the observed infrared spectrum of G21.5-0.9 is uncertain, we investigate two cases with (model 1) and without (model 2) infrared fitting.

\subsection{Model 1}
\begin{figure}
\plotone{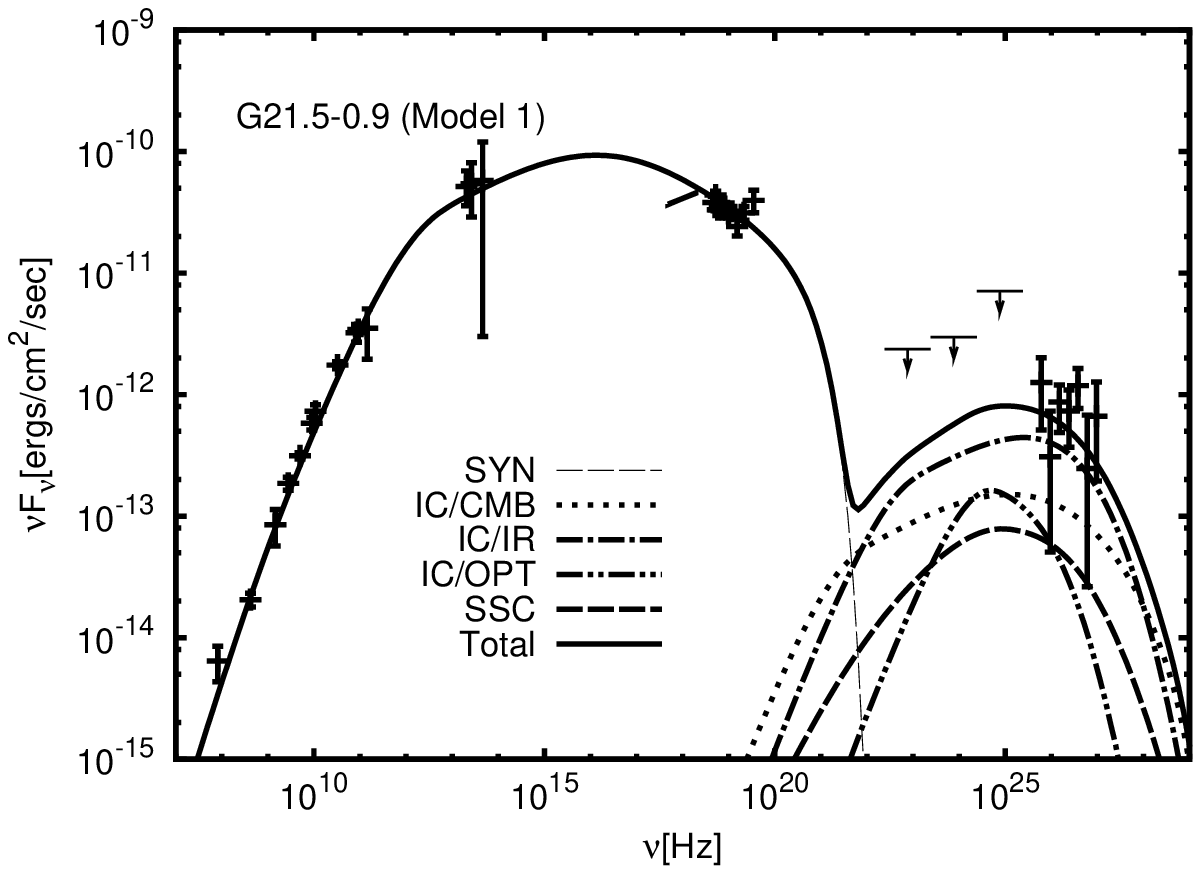}
\caption{Model spectrum of G21.5-0.9 at $t_{\rm age} = 1.0 \rm kyr$ for model 1, where fitting to the infrared observation is included.
The solid line is the total spectrum which is the sum of the synchrotron (thin-dashed line), IC/CMB (dotted line), IC/IR (dot-dashed line), IC/OPT (dot-dot-dashed line) and SSC (dashed line) spectra, respectively.
The observed data are taken from \citet{set89} (radio), \citet{gt98} (IR), \citet{tet10b, det09b} (X-ray), \citet{aet11, det08a} ($\gamma$-ray).
Used and obtained parameters are tabulated in Table \ref{tbl-1}.
\label{g21.5_model1_current}}
\end{figure}

Figure \ref{g21.5_model1_current} shows the model spectrum of G21.5-0.9 with the observational data including the infrared observation.
We fit the data with the parameters $\eta = 1.5 \times 10^{-2}$, $t_{\rm age} = 1.0 \rm kyr$, $\gamma_{\rm max} = 2.0 \times 10^9$, $\gamma_{\rm b} = 1.2 \times 10^5$, $\gamma_{\rm min} = 3.0 \times 10^3$, $p_1 = 1.0$, and $p_2 = 2.55$.
The fitted fraction parameter is three times larger than the Crab Nebula, but still much smaller than unity.
The fitted age $t_{\rm age} = 1.0 \rm kyr$ deviates from the characteristic age of the central pulsar $\tau_{\rm c} = 4.9 \rm kyr$.
Because the ratio of the fitted age to the characteristic age of the pulsar is described as $t_{\rm age} / \tau_{\rm c} = 2 / (n -1) \cdot [1 - (P_0 / P)^{(n-1)}]$, the current pulsar period is almost the same as the initial one ($P_0 / P \sim 0.9$).
The derived expansion velocity $v_{\rm PWN} = 980 \rm km / sec$ and the fitted age are consistent with the observed expansion rate \citep[][]{bb08}.
From the fitted age, we obtain a spin-down time $\tau_0 = 3.9 {\rm kyr}$ and an initial rotational energy of the pulsar $L_0 \cdot \tau_0 = 6.5 \times 10^{48} \rm erg$.
The current total energy injected into the PWN $E_{\rm tot}(1 \rm kyr) = 1.3 \times 10^{48} \rm erg$ is significantly below $L_0 \cdot \tau_0$, since $\tau_0 > t_{\rm age}$.
The current magnetic field strength of G21.5-0.9 turns out to be $B_{\rm now} = 64 \mu \rm G$. 
We choose $\gamma_{\rm min}$ and $\gamma_{\rm max}$ by taking $\nu_{\rm syn}(\gamma_{\rm min}, B_{\rm now}) \sim 10^{8} \rm Hz$ and $\nu_{\rm syn}(\gamma_{\rm max}, B_{\rm now})\sim 3 \times 10^{20} \rm Hz$, respectively.

There are two breaks in the calculated synchrotron spectrum in Figure \ref{g21.5_model1_current}.
One is the synchrotron cooling break frequency $\nu_{\rm c} \sim 3 \times 10^{15} \rm Hz$, which is determined from Equation (\ref{eq_break-frequency}), the other is the break in the particle injection $\nu_{\rm syn}(\gamma_{\rm b}, B_{\rm now}) \sim 8 \times 10^{11} \rm Hz$, which is much smaller than $\nu_{\rm c}$.
Since $2 < p_2 < 3$, the synchrotron spectrum in Figure \ref{g21.5_model1_current} is peaked at around $\nu_{\rm c}$.
Focusing on the calculated TeV spectrum in Figure \ref{g21.5_model1_current}, the IC/IR is found to dominate and the SSC contribution is negligible.
Although the energy density of the ISRF in optical band is twice as large as that in infrared band, Klein-Nishina effect significantly reduces the IC/OPT.
The calculated flux of the GeV $\gamma$-ray is almost an order of magnitude below the upper limit of the PWN component given by \citet{aet11}.
In hard X-rays, the observed spectrum in $5 \times 10^{18}$ -- $2 \times 10^{19} \rm Hz$ is well reproduced, while the pulsar emission appears to dominate above $2 \times 10^{19} \rm Hz$ \citep[][]{det09b}.
The observed spectrum in soft X-rays ($< 3 \times 10^{18} \rm Hz$) and the observed sharp break around $3 \times 10^{18} \rm Hz$ are difficult to reproduce in our model.
The observed soft X-ray spectrum is much harder than the calculated spectrum and such a hard spectrum does not smoothly connect to the observed infrared spectrum.
We will discuss about these discrepancies between the calculated and observed spectra in soft X-rays in Section \ref{dis_g21.5}.

\begin{figure}
\plottwo{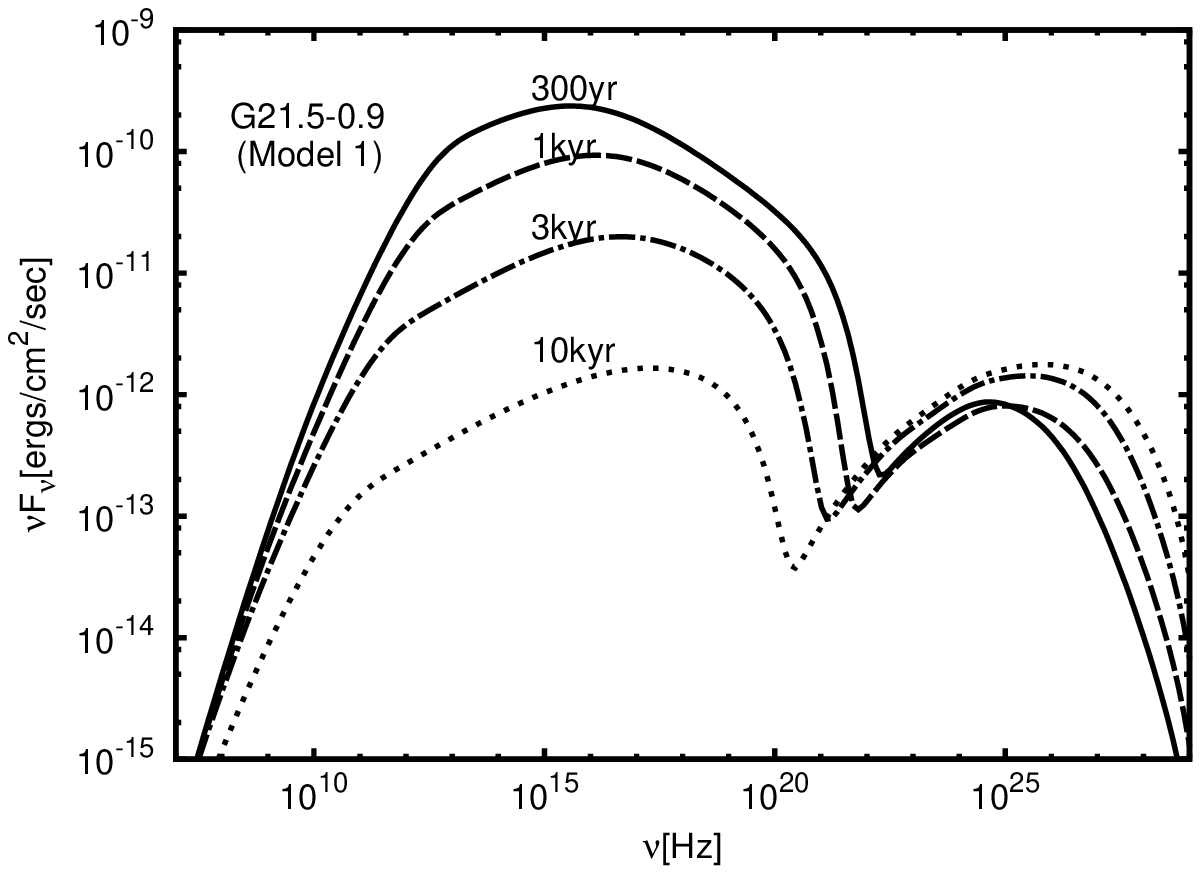}{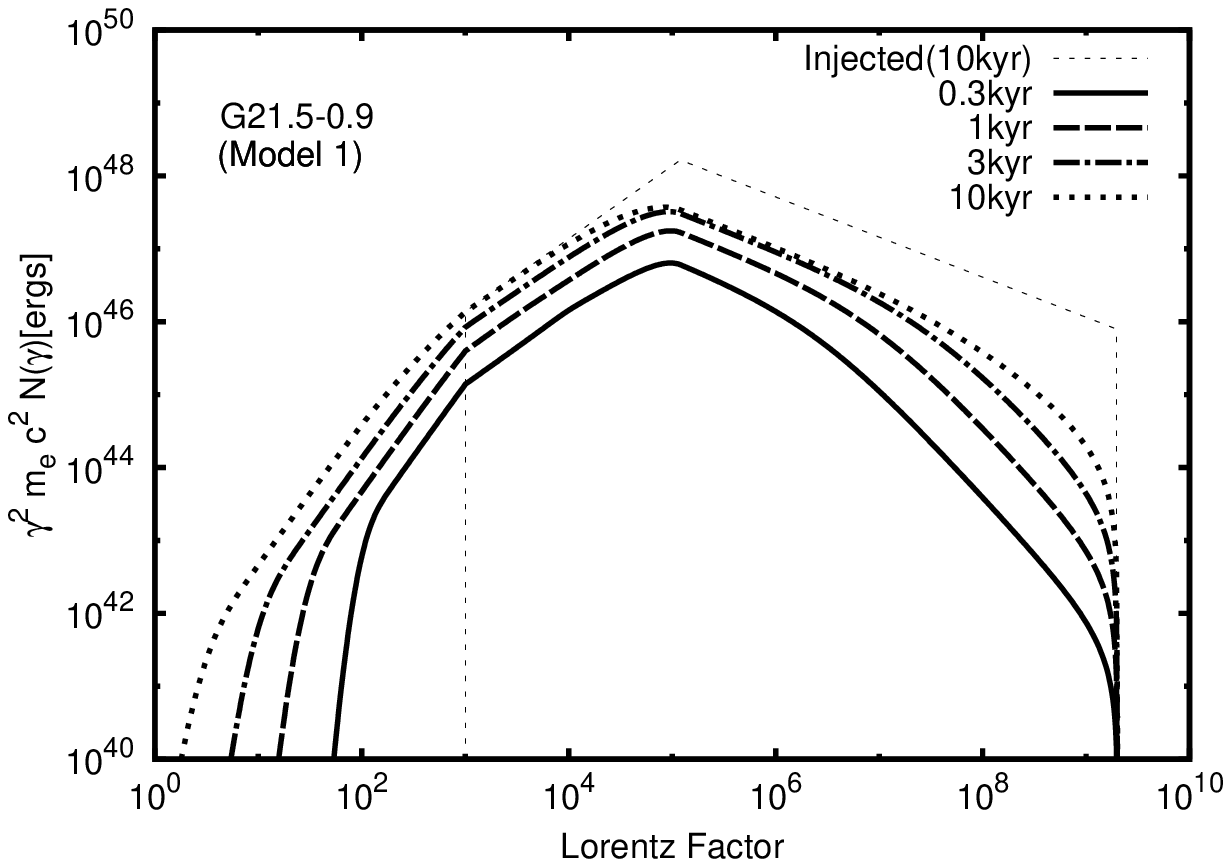}
\caption{Evolution of the emission spectrum (left panel) and the particle distribution (right panel) of G21.5-0.9 for model 1.
The solid line, dashed line, dot-dashed line, and dotted line correspond to 300yr, 1kyr, 3kyr, and 10kyr from birth, respectively.
The thin dotted line in the particle distribution (right panel) is the total injected particles at an age of 10kyr.
\label{g21.5_model1_evolution}}
\end{figure}

Figure \ref{g21.5_model1_evolution} shows the evolution of the emission spectrum (left panel) and that of the particle distribution (right panel) of G21.5-0.9 with the use of the same parameters as in Figure \ref{g21.5_model1_current}.
While the synchrotron flux decreases with time, the inverse Compton flux increases.
This feature is due to the decrease of the magnetic field strength with time and to the increase of the particle number as seen in the right panel.
Because the energy injection from the pulsar will continue till the time $t \sim \tau_0 = 3.9 \rm kyr$, the particles in the PWN increase (see the discussion in Section 3.2 of \citet{tt10} in details).
In the left panel, we can see the evolution of the synchrotron cooling break frequency $\nu_{\rm c}(t)$ (from $\sim 7 \times 10^{14} \rm Hz$ at 300 yr to $\sim 10^{17} \rm Hz$ at 10 kyr) as is predicted by Equation (\ref{eq_break-frequency}). 

\subsection{Model 2}
\begin{figure}
\plotone{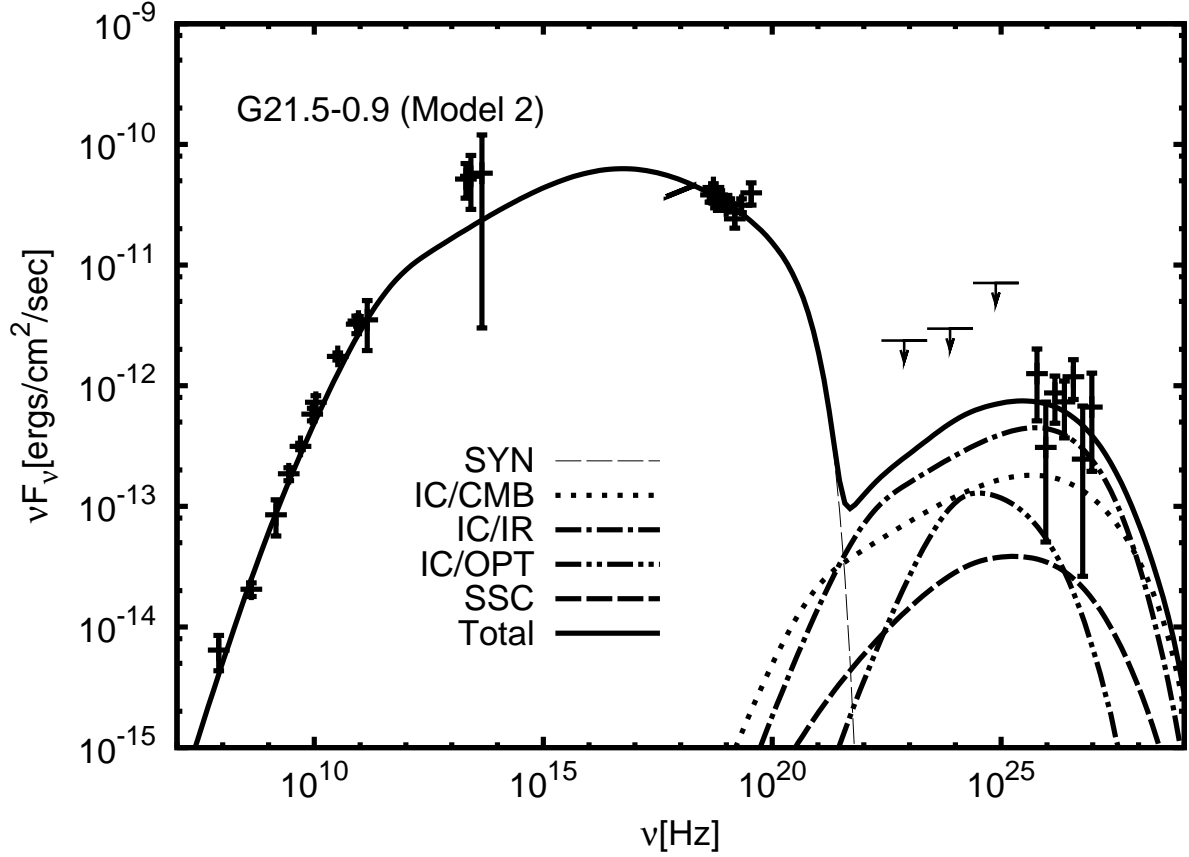}
\caption{Model spectrum of G21.5-0.9 at $t_{\rm age} = 1.0 \rm kyr$ for model 2, where we ignore the infrared observation \citep{gt98} in the spectral fitting.
Used and obtained parameters are tabulated in Table \ref{tbl-1}.
\label{g21.5_model2_current}}
\end{figure}

The calculated spectrum in model 1 does not reproduce the observed spectrum in infrared and soft X-rays at the same time.
Considering the uncertainties in the infrared observations, we try to fit the observed spectrum ignoring the infrared band.
Figure \ref{g21.5_model2_current} shows the model spectrum of G21.5-0.9 without infrared fitting.
All the fitted parameters are similar to model 1 (see Table \ref{tbl-1}), but a slightly smaller value of the fraction parameter $\eta = 8 \times 10^{-3}$ ($B_{\rm now} = 47 \mu \rm G$) is allowed.
Consequently, the synchrotron flux in infrared is smaller than model 1 and the synchrotron cooling break frequency $\nu_{\rm c} \sim 7 \times 10^{15} \rm Hz$ is a little larger than model 1.
The observed spectrum in soft X-rays and the observed sharp break at around $3 \times 10^{18} \rm Hz$ are not well reproduced, even if we ignore infrared observations.
The hard X-ray observation ($5 \times 10^{18}$ -- $2 \times 10^{19} \rm Hz$) is reproduced as well as model 1.

\subsection{Discussion}\label{dis_g21.5}
Both model 1 and 2 reproduce the observational data reasonably well.
The fraction parameter $\eta$, i.e., the magnetic field strength $B_{\rm now}$, is the only notable difference between the models.
Future observations in infrared and optical bands will provide more accurate spectrum  of G21.5-0.9 and could distinguish the models.
As for the other parameters, the values of $\gamma_{\rm b}$ and $p_1$ are smaller than the Crab Nebula in both models.
However, the high energy power-law index at injection $p_2$ is very similar to the Crab Nebula.

We compare our results with the model by \citet{det09a}.
They solved hydrodynamic equations and the evolution of the magnetic field is separately calculated by the induction equation.
The evolution of the broad band spectrum is calculated with a one-zone approach with the use of the volume averaged magnetic field.
Their adopted energy density of the ISRF in infrared band $U_{\rm IR} = 1.0 \rm eV/cm^3$ and $t_{\rm age} = 1 \rm kyr$ are the same as ours.
They obtain a particle injection distribution with $p_1 = 1.0$, $p_2 = 2.6$ and $\gamma_{\rm b} \sim 8 \times 10^4$, which are almost the same as ours.
Their obtained current magnetic field strength $\sim 24 \mu \rm G$ is smaller than ours, $B_{\rm now} = 64 \mu \rm G$ and $47 \mu \rm G$ for model 1 and 2, respectively.
Their results are basically consistent with ours.

Lastly, we discuss the discrepancy between our model spectra and the observations in soft X-rays.
\citet{tet10b} show that there is an observed spectral break between the soft and hard X-rays at $\sim 3 \times 10^{18} \rm Hz$.
One might consider that this spectral break corresponds to the synchrotron cooling break frequency $\nu_{\rm c}$.
With $t_{\rm age} = 1 \rm kyr$, the value of $\nu_{\rm c} \sim 3 \times 10^{18} \rm Hz$ requires the magnetic field strength to be $\sim 6 \mu \rm G$ from Equation (\ref{eq_break-frequency}).
However, the current magnetic field of $\sim 6 \mu \rm G$ seems unlikely because the observed $\gamma$-rays to X-rays flux ratio would demands the local ISRF energy density around G21.5-0.9 to be much smaller than the contribution from the CMB.
Note that it is hard to reproduce the observed sharp spectral break at $\sim 3 \times 10^{18} \rm Hz$, even if the magnetic field strength is $\sim 6 \mu \rm G$.
This is because the evolution of the magnetic field $B(t)$ and the particle injection $Q_{\rm inj}(\gamma, t)$ makes the synchrotron cooling break rather smooth.
We thus conclude that any one-zone spectral evolution models fail to reproduce the current soft X-ray observation.
One possible resolution is a modification of the observed spectrum in soft X-rays with relatively large interstellar extinction toward G21.5-0.9 as discussed by \citet{set01}.
Another possibility is the effect of spatial variation of the soft X-ray photon index $\Gamma_X$ as discussed by \citet{set00}.

\section{G54.1+0.3}\label{g54.1}
G54.1+0.3 is a center filled or possibly a composite SNR and its PWN component is observed in radio \citep[][]{g85, vet86, vb88, let10a}, X-rays \citep[][]{let01}, and TeV $\gamma$-rays \citep[][]{aet10}.
In X-rays, a jet-torus structure is observed similarly to the Crab Nebula \citep[][]{let02}.
The size of the radio PWN $\sim 2.5' \times 2.0'$ is comparable with that in X-rays \citep[][]{let10a, bet10b}.
Recently, diffuse X-ray emission surrounding the PWN was detected by \citet{bet10b}, which is a possible counterpart of SN ejecta component, while a possible SNR shell is observed in radio \citep[][]{let10a}.
The region around G54.1+0.3 has been observed in infrared \citep[][]{ket08, tet10a}.
Although no infrared counterpart of G54.1+0.3 PWN is observed, there are several bright sources around G54.1+0.3.
The central pulsar of G54.1+0.3 (PSR J1930+1852) is observed in radio and X-rays with a period $P = 1.36 \times 10^{-1} \rm sec$, its time derivative $\dot{P} = 7.51 \times 10^{-13} \rm sec \cdot sec^{-1}$ ($\tau_c = 2.9 \rm kyr$) and unknown braking index \citep[][]{cet02}.
So that we assume $n = 3$.
The actual age of G54.1+0.3 is unknown.
We assume a distance to G54.1+0.3 of 6.2kpc \citep[][]{let08} and then we assume that G54.1+0.3 PWN is a sphere of radius 1.8pc at $(r, z) \sim (7.5 {\rm kpc}, 0 {\rm pc})$.
We get $(U_{\rm IR}, U_{\rm OPT}) = (0.5 {\rm eV/cm^3}, 0.5 {\rm eV/cm^3})$ for the energy density of the ISRF in model 1.
We also investigate in model 2 the probability that the observed infrared sources around G54.1+0.3 significantly contribute to the local ISRF.

\subsection{Model 1}
\begin{figure}
\plotone{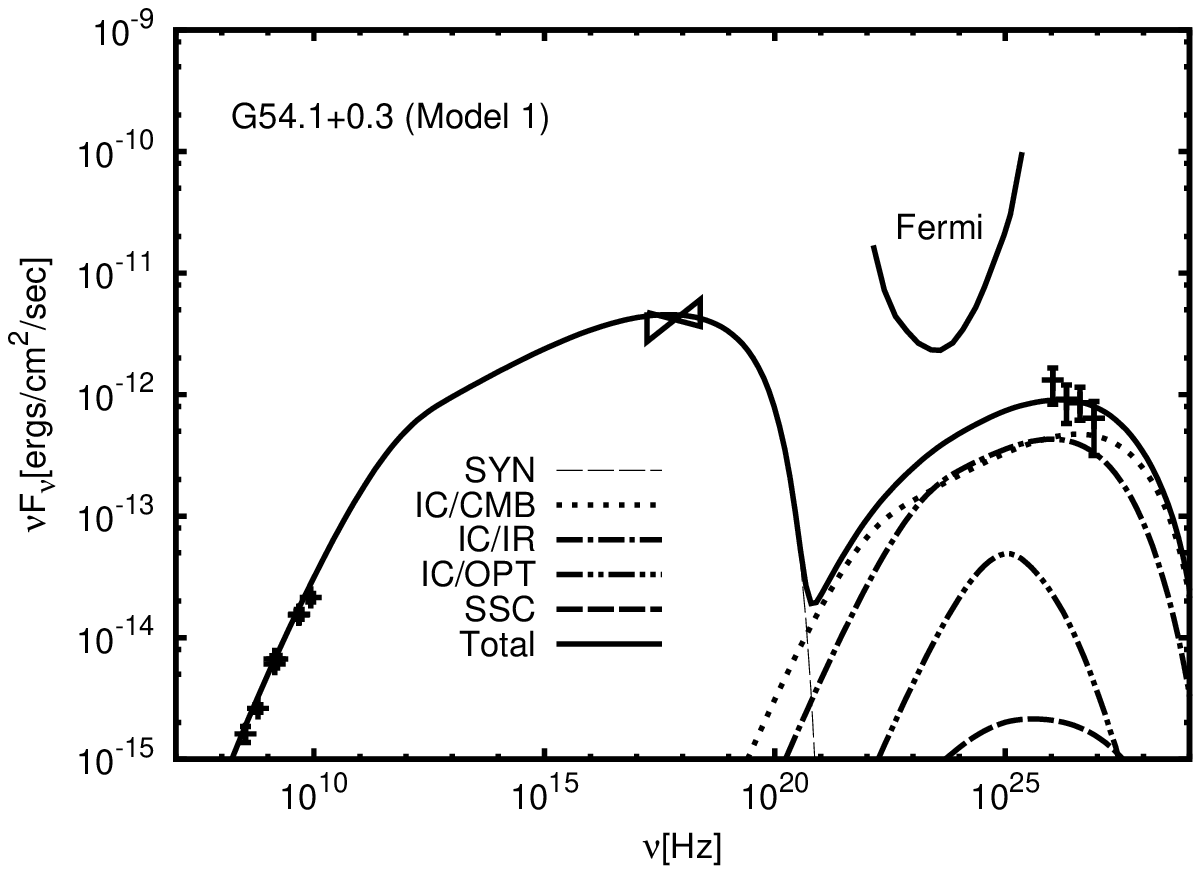}
\caption{Model spectrum of G54.1+0.3 at $t_{\rm age} = 2.3 \rm kyr$ for model 1.
The observational data and the 1yr, 5$\sigma$ sensitivity for {\it Fermi} LAT are plotted.
This sensitivity is very similar to the upper limit obtained for G21.5-0.9 shown in Figure \ref{g21.5_model1_current}.
The observed data are taken from \citet{g85, vet86, vb88, let10a} (radio), \citet{let01} (X-ray), \citet{aet10} ($\gamma$-ray).
Used and obtained parameters are tabulated in Table \ref{tbl-1}.
\label{g54.1_model1_current}}
\end{figure}

Figure \ref{g54.1_model1_current} shows the model spectrum of G54.1+0.3 for model 1 with the observational data.
We can reproduce the observed non-thermal spectrum with the parameters $\eta = 3.0 \times 10^{-4}$, $t_{\rm age} = 2.3 \rm kyr$, $\gamma_{\rm max} = 1.0 \times 10^9$, $\gamma_{\rm b} = 3.0 \times 10^5$, $\gamma_{\rm min} = 2.0 \times 10^4$, $p_1 = 1.2$, and $p_2 = 2.55$.
The fraction parameter is 0.06 times the Crab Nebula's value.
The fitted age $t_{\rm age} = 2.3 \rm kyr$ is comparable with the characteristic age of the central pulsar $\tau_c = 2.9 \rm kyr$, and the current period is almost twice as large as the initial period.
The corresponding expansion velocity $v_{\rm PWN} = 770 \rm km / sec$ is less than a half of that of the Crab Nebula 1800 km/sec.
We obtain a spin-down time $\tau_0 = 0.6 \rm{kyr} < t_{\rm age}$ and an initial rotational energy  $L_0 \cdot \tau_0 = 5.4 \times 10^{48} \rm erg$.
The current total energy injected into the PWN $E_{\rm tot}(2.3 \rm kyr) = 4.3 \times 10^{48} \rm erg$ is close to the value of $L_0 \cdot \tau_0$.
The current magnetic field strength of G54.1+0.3 turns out to be $B_{\rm now} = 6.7 \mu \rm G$, which is much smaller than that of the Crab Nebula and G21.5-0.9.

The synchrotron cooling break frequency $\nu_{\rm c} \sim 5 \times 10^{17} \rm Hz$ is  much larger than the characteristic synchrotron frequency corresponding to the break energy $\nu_{\rm syn}(\gamma_{\rm b}, B_{\rm now}) \sim 7 \times 10^{11} \rm Hz$.
In $\gamma$-rays, the ISRF energy density in infrared band is twice as large as that of the CMB, but the contributions of the IC/CMB and the IC/IR are comparable because the IC/IR is in mildly Klein-Nishina regime.
Going into details, the observed X-ray spectrum seems consistent with the model spectrum, while the observed $\gamma$-ray spectrum seems softer.
Because particles with almost the same energy contribute both to the observed emission in X-rays (SYN) and $\gamma$-rays (IC/CMB in Thomson regime), it is difficult to fit the spectral slopes in both frequency ranges at the same time.

\begin{figure}
\plottwo{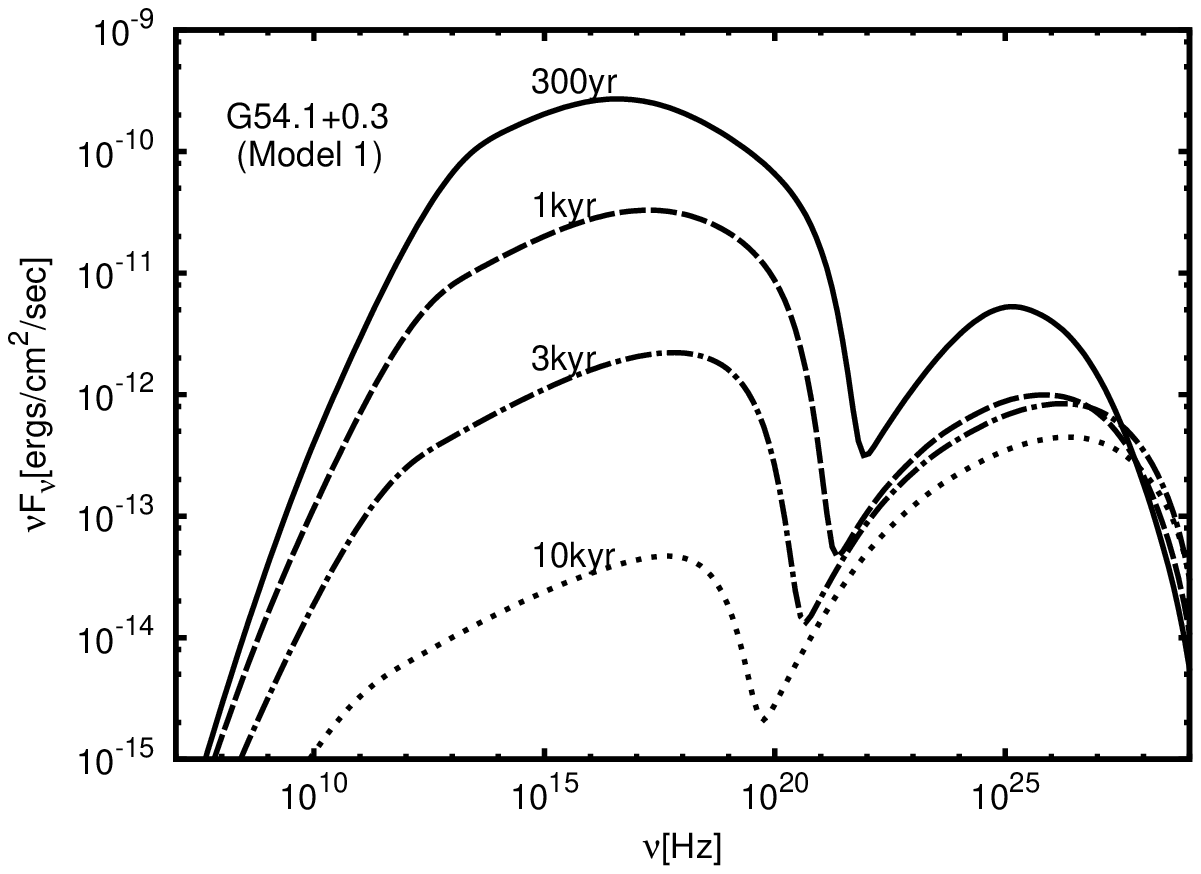}{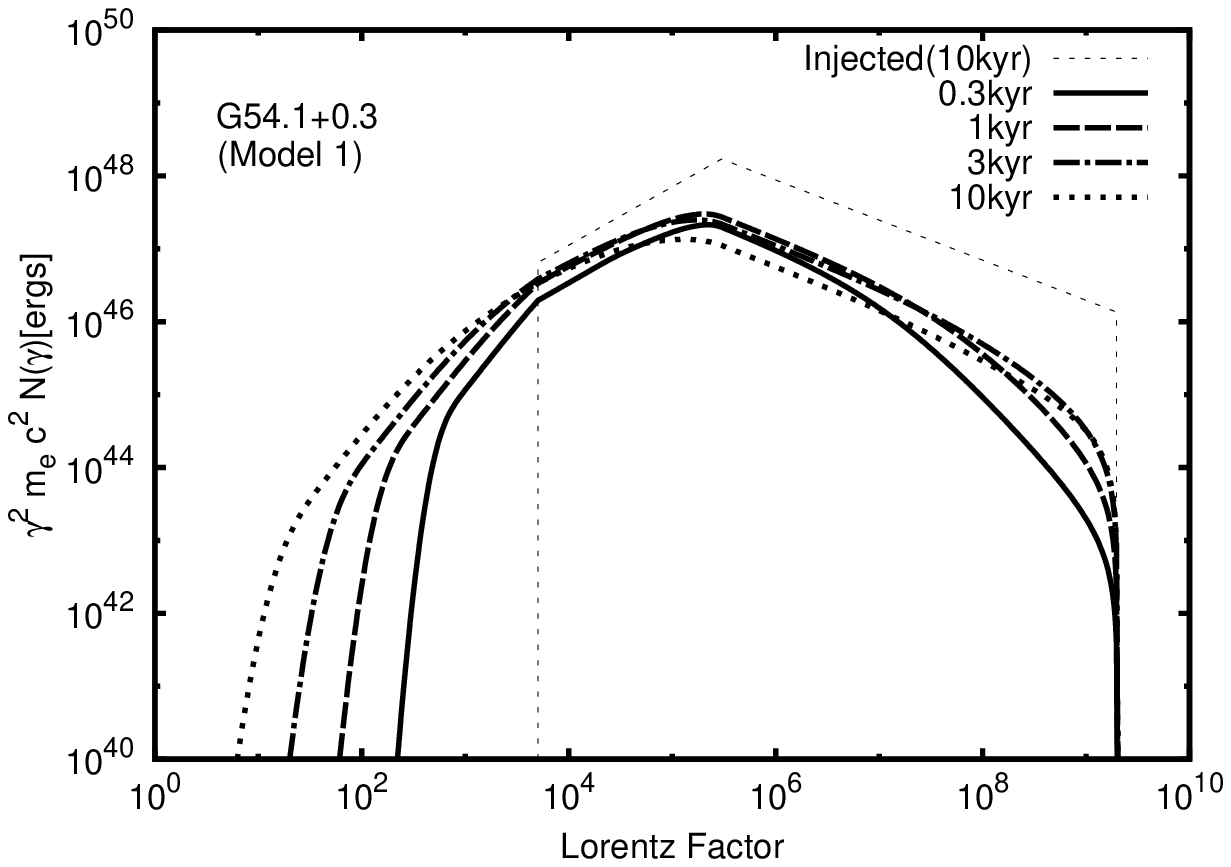}
\caption{Evolution of the emission spectrum (left panel) and the particle distribution (right panel) of G54.1+0.3 for model 1.
\label{g54.1_model1_evolution}}
\end{figure}

Figure \ref{g54.1_model1_evolution} shows the evolution of the emission spectrum (left panel) and the particle distribution (right panel) of G54.1+0.3 with the use of the same parameters as in Figure \ref{g54.1_model1_current}.
Both the synchrotron flux and the IC flux decrease with time, but the IC flux deceases more slowly.
This feature is because the decrease of the magnetic field strength is much faster than the particle distribution while the ISRF energy density is constant.
This behavior of the particle distribution has already been discussed in Section 3.2 of \citet{tt10}.
In short, the evolution of the particle number is approximately expressed as $N(\gamma, t) \sim Q(\gamma, t) \cdot t \cdot \tau_{\rm cool}(\gamma, t) / (t + \tau_{\rm cool}(\gamma, t))$ and the injection from the pulsar decreases as $Q(\gamma, t) \propto t^{-2}$ after the time $t \sim \tau_0 = 600 \rm yr$.
In synchrotron cooling regime ($\tau_{\rm cool} = \tau_{\rm syn} < \tau_{\rm ad} \sim t$), the particle number increases as $N \propto t$, while in adiabatic cooling regime ($\tau_{\rm cool} = \tau_{\rm ad}$), the particle number decreases as $N \propto t^{-1}$, where we use $\tau_{\rm syn} \propto t^3$ for $t > \tau_0$.
The left panel of Figure \ref{g54.1_model1_evolution} shows that the synchrotron cooling break frequency $\nu_{\rm c}(t)$ will continue to increase with time till an age of 3kyr (Equation (\ref{eq_break-frequency})).
Because $\gamma_{\rm c}(\rm 10kyr) > \gamma_{\rm max}$, the synchrotron spectrum will be peaked at $\nu_{\rm syn}(\gamma_{\rm max})$, not $\nu_{\rm c}$, at an age of 10kyr. 

\subsection{Model 2}
\begin{figure}
\plotone{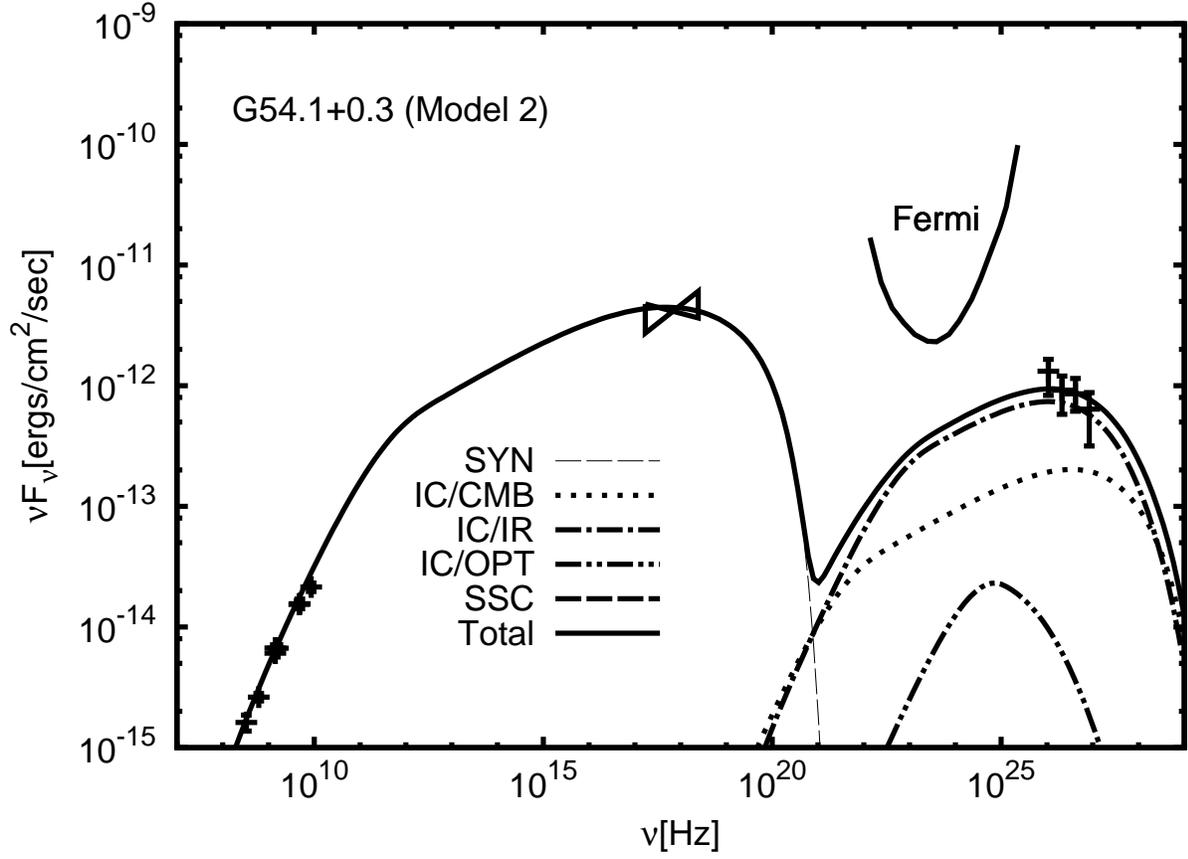}
\caption{Model spectrum of G54.1+0.3 at $t_{\rm age} = 1.7 \rm kyr$ for model 2, where an enhanced local ISRF is assumed.
{\it Fermi} LAT sensitivity and the observed data are the same as in Figure \ref{g54.1_model1_current}.
Used and obtained parameters are tabulated in Table \ref{tbl-1}.
\label{g54.1_model2_current}}
\end{figure}

The region around G54.1+0.3 has been observed in infrared \citep[][]{ket08, tet10a} and observations suggest that the ISRF around G54.1+0.3 could be larger than that of the average values of the Galaxy adopted in model 1.
Figure \ref{g54.1_model2_current} shows the model spectrum of G54.1+0.3 for model 2 when the energy density of the ISRF in infrared band has been chosen to be four times larger than model 1 ($U_{\rm IR} = 2.0 \rm eV/cm^3$).
All the fitted parameters are similar to model 1 except for the fraction parameter and the fitted age; they are $\eta = 2.0 \times 10^{-3}$ and $t_{\rm age} = 1.7 \rm kyr$, respectively (see Table \ref{tbl-1}).
The fitted fraction parameter is close to the values of the Crab Nebula.
The fitted age $t_{\rm age} = 1.7 \rm kyr$ is similar to the independent estimate of 1.5 kyr from the dynamical interaction with SNR by \citet{c05}.
Because the fitted age $t_{\rm age}$ changes from model 1, the derived parameters change also from model 1, $v_{\rm PWN} = 1040 \rm km / sec$, $\tau_0 = 1.2 \rm kyr$ and $L_0 \cdot \tau_0 = 2.6 \times 10^{48} \rm erg$ ($E_{\rm tot}(1.7 \rm kyr) = 1.6 \times 10^{48} \rm erg$), respectively.
The current magnetic field strength $B_{\rm now} = 10 \mu \rm G$ becomes a little larger than model 1.
We can see the trend that a larger energy density of the ISRF leads to a larger $\eta$ and $B_{\rm now}$ and a smaller value of $E_{\rm tot}(t_{\rm age})$ and $t_{\rm age}$.
This trend can be understood following a similar discussion made in Section 3.3 of \citet{tt10} (see Section \ref{discussion}).

\subsection{Discussion}
We favor model 2 because the observed $\gamma$-ray spectrum is somewhat better reproduced by model 2, although both model 1 and 2 can reproduce the observational data reasonably well; model 2 has a softer $\gamma$-ray spectrum than model 1 through the contribution of the IC/IR is larger than the IC/CMB.
Because the difference of the ISRF energy density in the models appears at the frequency where the Klein-Nishina effect works, future observations in higher energy $\gamma$-rays (1 -- 100TeV), such as CTA, would provide better information on the correct values of the ISRF.
In contrast to $\eta$ and $t_{\rm age}$, the parameters of the particle injection for both models are similar.
The low energy power-law index at injection $p_1$ is different from the Crab Nebula and G21.5-0.9, but the high energy power-law index at injection $p_2$ is very similar.

The spectral evolution of G54.1+0.3 was also studied by \citet{let10b}.
In addition to pure-lepton model \citep[Figure 2 of ][]{let10b}, they studied a lepton-hadron hybrid model of the broad band spectrum \citep[Figure 3 of ][]{let10b}.
In the pure-lepton model, they obtained the parameters $p_1 = 1.2$ and $\gamma_{\rm b} = 5 \times 10^5$, which are similar to ours.
However, their obtained $p_2 = 2.8$ and adopted $U_{\rm IR} \sim 3.3 \rm eV/cm^3$ are different from ours in model 2, $p_2 = 2.55$ and $U_{\rm IR} \sim 2.0 \rm eV/cm^3$, respectively.
Difference appears in the X-ray spectrum, which in their model seems softer than ours.
The current magnetic field strength in their model $\sim 10 \mu \rm G$ is the same as our model 2.
We conclude that their pure-lepton model is almost consistent with model 2 of ours except for the X-ray spectrum.

\citet{let10b} argued that the lepton-hadron hybrid model ($B \sim 80 \mu \rm G$) is better than the pure-lepton model because the current magnetic field $B \sim 10 \mu \rm G$ in pure-lepton model is much weaker than the observational indication by \citet{let10a}.
\citet{let10a} estimated an equipartition magnetic field of 38 $\mu \rm G$ from the radio luminosity of PWN and a magnetic field of 80 - 200 $\mu \rm G$ from the lifetime of X-ray emitting particles.
We consider that these estimates by \citet{let10a} are not robust.
Generally, the magnetic field strength is sub-equipartition for all PWNe we studied, i.e., $\eta << 1$, so that $B = 10 \mu \rm G$ is just a reasonably expected value for this PWN.
\citet{let10b} also argued that the observed $\gamma$-ray photon index is better fitted by the lepton-hadron hybrid model than the pure-lepton model.
However, the calculated $\gamma$-ray spectrum in the pure-lepton model changes with local ISRF energy density and temperature.
We believe that leptonic model with small magnetic field is consistent with the current observations.

\section{Kesteven 75}\label{kes75}
Kes 75 is a composite SNR and its PWN component is observed in radio \citep[][]{set89, bg05}, X-rays \citep[][]{het03, met08}, and TeV $\gamma$-rays \citep[][]{det07a}.
In X-rays, a jet-torus structure is observed \citep[][]{net08}.
This jet-torus structure is surrounded by a diffuse X-ray nebula, which has almost the same extent of the radio PWN $\sim 26'' \times 20''$ \citep[][]{het03}.
A part of the SNR shell is observed in radio and X-rays \citep[][]{bg05, het03}.
The central pulsar of Kes 75 (PSR J1846-0258) is observed in X-rays, with a period $P = 3.26 \times 10^{-1} \rm sec$, its time derivative $\dot{P} = 7.08 \times 10^{-12} \rm sec \cdot sec^{-1}$ and braking index $n = 2.65 \pm 0.01$ \citep[][]{let06}.
Although the characteristic age of the central pulsar $\tau_{\rm c} = 0.7 \rm kyr$ suggests it is very young, the actual age of the Kes 75 is unknown.
PSR J1846-0258 has a large surface dipole magnetic field and a magnetar-like burst was observed in 2006 \citep[][]{get08}.
No radio counterpart of PSR J1846-0258 is observed, common to most of the magnetar candidates \citep[][]{aet08}.
We consider two cases for the distance to Kes 75: 6kpc \citep[][]{lt08} and 10.6kpc \citep[][]{set09} as model 1 and 2, respectively.
We assume that Kes 75 PWN is a sphere of radius 0.29pc at $(r, z) \sim (4 {\rm kpc}, 30 {\rm pc})$ or 0.5pc at $(r, z) \sim (5 {\rm kpc}, 50 {\rm pc})$ for model 1 and 2, respectively.
The energy density of the ISRF is $(U_{\rm IR}, U_{\rm OPT}) = (1.2 {\rm eV/cm^3}, 2.0 {\rm eV/cm^3})$ for model 1 and $(U_{\rm IR}, U_{\rm OPT}) = (1.0 {\rm eV/cm^3}, 2.0 {\rm eV/cm^3})$ for model 2, respectively.

\subsection{Model 1}
\begin{figure}
\plotone{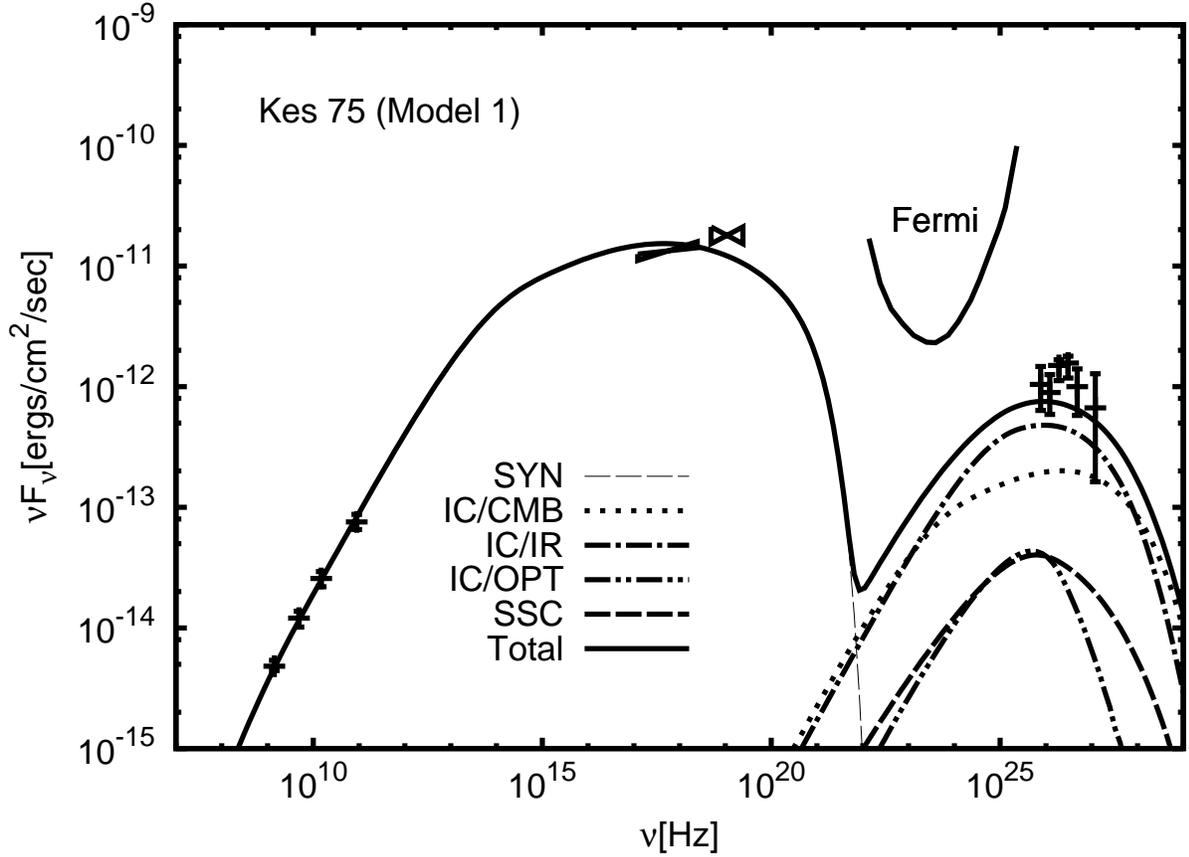}
\caption{Model spectrum of Kes 75 at $t_{\rm age} = 0.7 \rm kyr$ for model 1, where the distance is taken to be 6kpc.
The observational data and the 1yr, 5$\sigma$ sensitivity for {\it Fermi} LAT are plotted.
The observed data are taken from \citet{set89, bg05} (radio), \citet{het03, met07} (X-ray), \citet{det07a} ($\gamma$-ray).
Used and obtained parameters are tabulated in Table \ref{tbl-1}.
\label{kes75_model1_current}}
\end{figure}

Figure \ref{kes75_model1_current} shows the model spectrum of Kes 75 when the distance to Kes 75 is assumed to be 6kpc together with observational data.
We fit the data with the parameters $\eta = 5.0 \times 10^{-5}$, $t_{\rm age} = 0.7 \rm kyr$, $\gamma_{\rm max} = 2.0 \times 10^9$, $\gamma_{\rm b} = 2.0 \times 10^6$, $\gamma_{\rm min} = 5.0 \times 10^3$, $p_1 = 1.6$, and $p_2 = 2.5$.
The fraction parameter is very small and two orders of magnitude smaller than the Crab Nebula.
The fitted age $t_{\rm age} = 0.7 \rm kyr$ is very close to the characteristic age $\tau_{\rm c} = 0.7 \rm kyr$ and the expansion velocity $v_{\rm PWN} = 420 \rm km/sec$ is rather slow.
The pulsar parameters $\tau_0 = 0.2 \rm kyr$ and $L_0 \cdot \tau_0 = 1.5 \times 10^{48} \rm erg$ are also smaller than those of the other PWNe.
The small $\tau_0$ may be related to the large magnetic field of the PSR J1846-0258.
The current total energy $E_{\rm tot}(0.7 \rm kyr) = 9.1 \times 10^{47} \rm erg$ is roughly half of $L_0 \cdot \tau_0$.
The current magnetic field strength of Kes 75 turns out to be $B_{\rm now} = 20 \mu \rm G$.
Despite $\eta$ being more than an order of magnitude smaller than in the other PWNe, $B_{\rm now}$ is not so different because the size of the Kes 75 PWN is small, and accordingly, $v_{\rm PWN}$ is small.

Focusing on the detailed spectral features, the observed flux in hard X-rays is not well reproduced and a few times larger than the model prediction although the characteristic synchrotron frequency corresponding to $\gamma_{\rm max}$ extends to $\sim 3 \times 10^{20} \rm Hz$.
The hard X-ray observation is difficult to reproduce because the synchrotron cooling break frequency $\nu_{\rm c} \sim 2 \times 10^{17} \rm Hz$ is located at the soft X-rays (see discussion in Section \ref{dis_kes75}).
The $\gamma$-ray emission is the IC/IR dominant like G21.5-0.9 and G54.1+0.3.

\begin{figure}
\plottwo{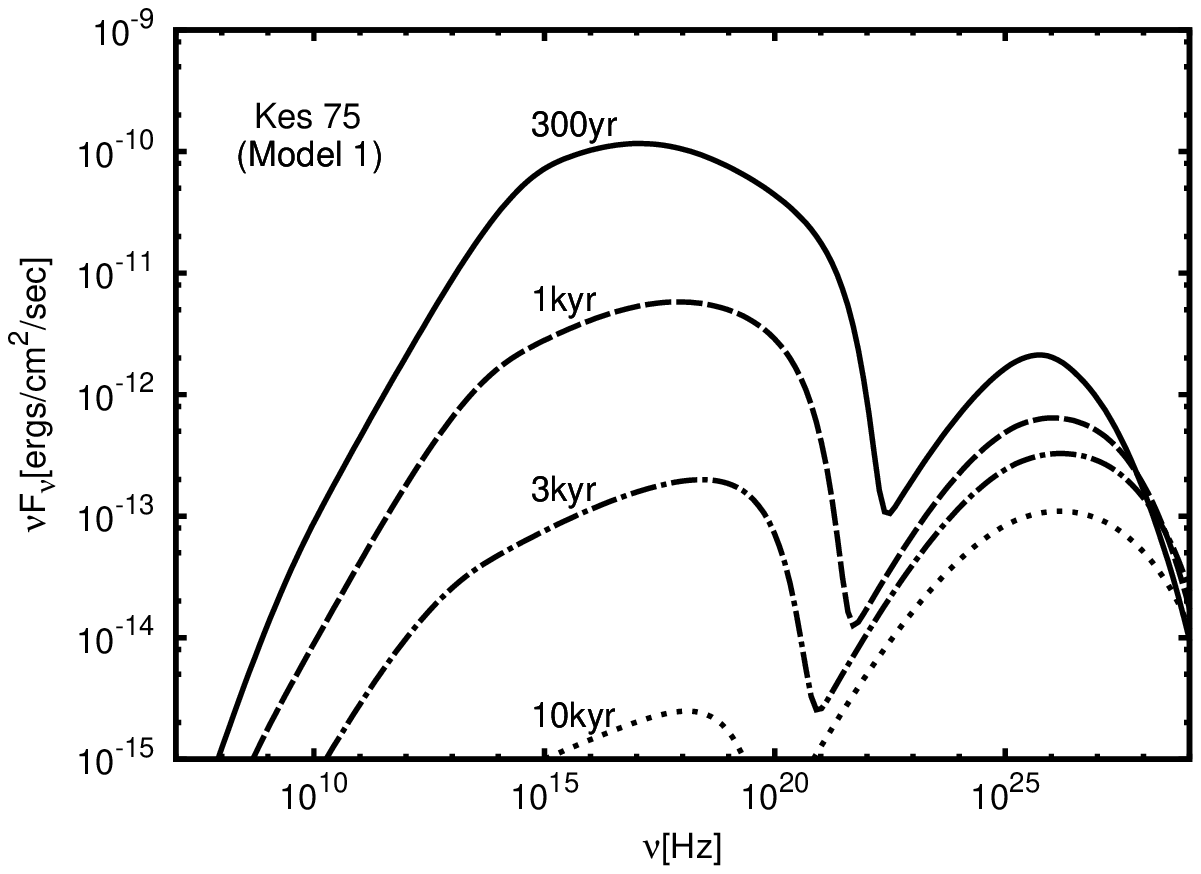}{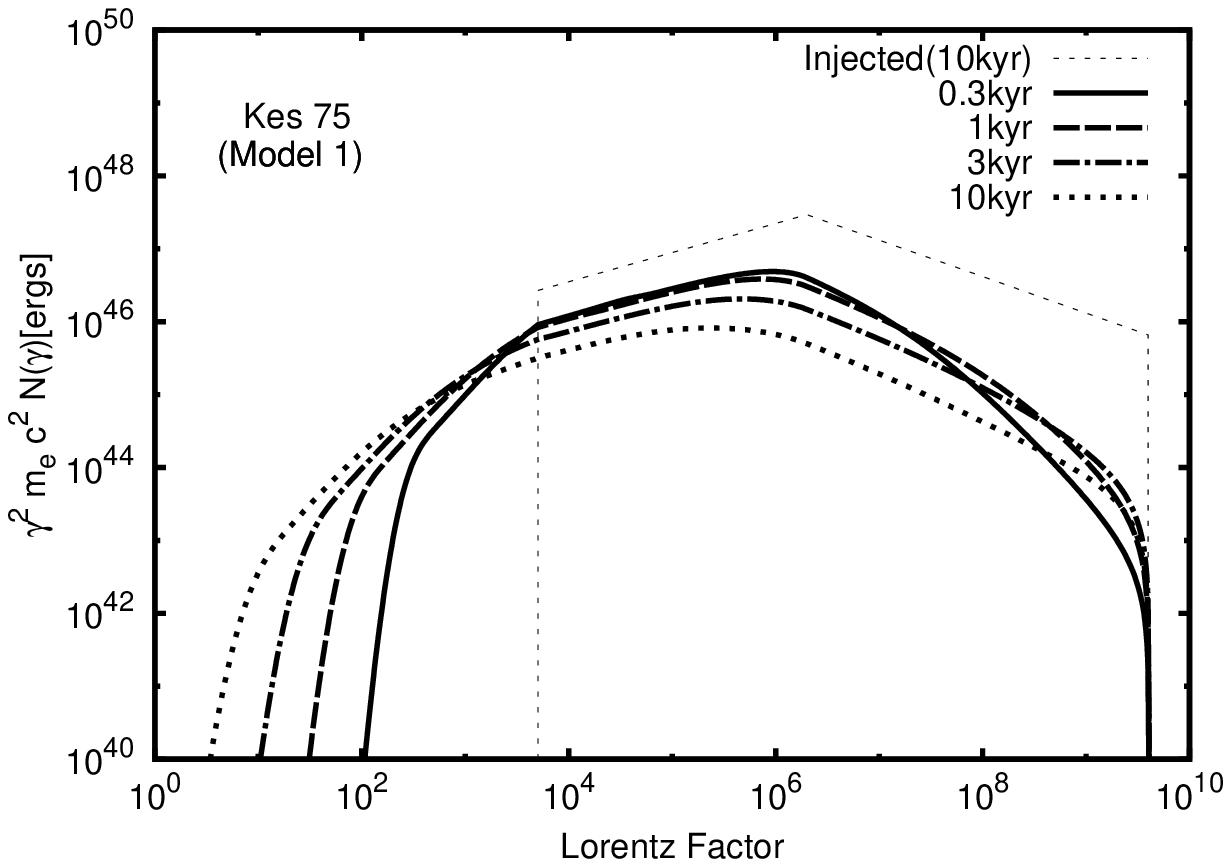}
\caption{Evolution of the emission spectrum (left panel) and the particle distribution (right panel) of Kes 75 for model 1.
\label{kes75_model1_evolution}}
\end{figure}

Figure \ref{kes75_model1_evolution} shows the evolution of the emission spectrum (left panel) and the particle distribution (right panel) of Kes 75 with the use of the same parameters as in Figure \ref{kes75_model1_current}.
Both the synchrotron flux and the IC flux decrease with time.
This feature is similar to G54.1+0.3 (Figure \ref{g54.1_model1_evolution}) because of the small spin-down time $\tau_0 = 0.2 \rm kyr$.

\subsection{Model 2}
\begin{figure}
\plotone{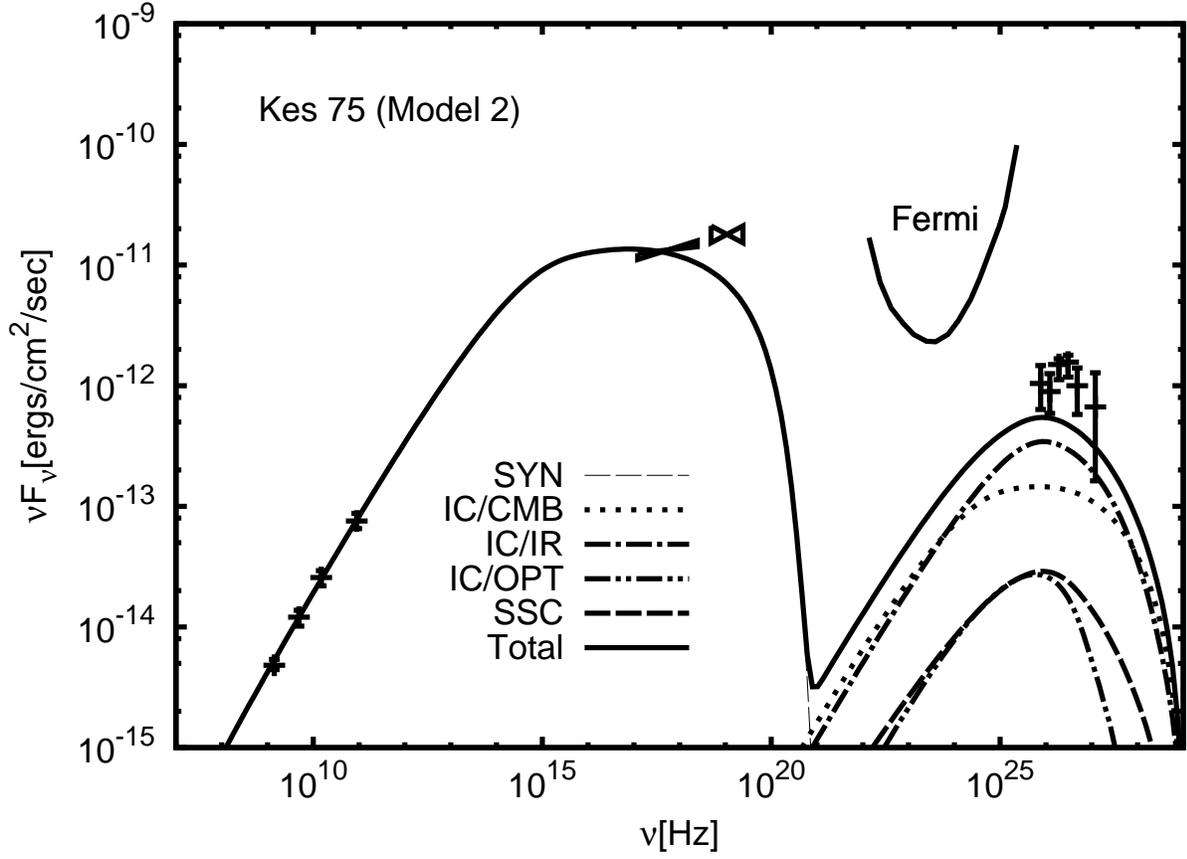}
\caption{Model spectrum of Kes 75 at $t_{\rm age} = 0.88 \rm kyr$ for model 2, where the distance is taken to be 10.6kpc.
{\it Fermi} LAT sensitivity and the observed data are the same as in Figure \ref{kes75_model1_current}.
Used and obtained parameters are tabulated in Table \ref{tbl-1}.
\label{kes75_model2_current}}
\end{figure}

Figure \ref{kes75_model2_current} shows the model spectrum of Kes 75 when the distance to Kes 75 is assumed to be 10.6kpc together with observational data.
We calculate the spectrum with the parameters $\eta = 6.0 \times 10^{-6}$, $t_{\rm age} = 0.88 \rm kyr$, $\gamma_{\rm max} = 1.0 \times 10^9$, $\gamma_{\rm b} = 5.0 \times 10^6$, $\gamma_{\rm min} = 5.0 \times 10^3$, $p_1 = 1.4$, and $p_2 = 2.5$, but the calculated $\gamma$-ray flux is a few times smaller than the observed one.
The fraction parameter is an order of magnitude smaller than model 1.
The fitted age $t_{\rm age} = 0.88 \rm kyr$ is almost the maximum value given by Equation (\ref{eq_age_tc_t0}).
The current magnetic field strength $B_{\rm now}$ is $24 \mu \rm G$ similar to model 1 because the adopted ISRF energy density is similar.
On the other hand, $E_{\rm tot} = 1.7 \times 10^{50} \rm erg$ is more than two orders of magnitude larger than model 1 given the larger $\gamma$-ray luminosity.
The pulsar parameters $\tau_0 = 3 \rm yr$ and $L_0 \cdot \tau_0 = 2.1 \times 10^{50} \rm erg$ $\sim E_{\rm tot}$ are more extreme than model 1.

To better reproduce the observed $\gamma$-ray flux in model 2, $E_{\rm tot}$ larger than $1.7 \times 10^{50} \rm erg$ is required with the adopted ISRF energy density.
$E_{\rm tot} > 1.7 \times 10^{50} \rm erg$ leads to $\tau_0 < 3 \rm yr$ and $L_0 \cdot \tau_0 > 2.1 \times 10^{50} \rm erg$.
However, $\tau_0 = 3 \rm yr$ is extremely short compared with $t_{\rm age} = 0.88 \rm kyr$ and $L_0 \cdot \tau_0 = 2.1 \times 10^{50} \rm erg$ is three times larger than the Crab Pulsar.
On the other hand, these extreme $\tau_0$, $L_0 \cdot \tau_0$ and also $\eta$ may be allowed given the unique properties of the central pulsar PSR J1846-0258, and in this case an energy density of the ISRF a few times larger would reproduce the observed $\gamma$-ray flux.

\subsection{Discussion}\label{dis_kes75}
We favor model 1 rather than model 2 when we compare the parameters with the other PWNe, because model 2 of Kes 75 is clearly more extreme than model 1.
However, we find Kes 75 peculiar in its results, even for the distance of 6kpc (model 1). 
The fitted values of $\eta = 5 \times 10^{-5}$ and $v_{\rm PWN} = 420 \rm km/sec$ are significantly smaller than other PWNe we studied so far.
The parameters $\eta$ and $v_{\rm PWN}$ become large, if we increase the local ISRF energy density $U_{\rm ISRF}$ in the same manner as for model 2 of G54.1+0.3.
However, to get two orders of magnitude larger value of $\eta$ and $v_{\rm PWN} \sim 1000 \rm km/sec$, more than an order magnitude larger $U_{\rm ISRF}$ is needed (see Section \ref{discussion} in details) and the age of the pulsar $t_{\rm age}$ becomes around 300 yr, which is as small as the age of SNR Cassiopeia A.
Because Kes 75 is the youngest PWN in our study, more precise studies of how a PWN inside a SNR is created may be important besides the magnetar-like properties of its central pulsar PSR J1846-0258.
As for the particle injection, the parameters are not unusual except for a little larger value of $\gamma_{\rm b}$ than other PWNe.
Especially, the high energy power-law index at injection $p_2 = 2.5$ is similar to other PWNe.

\citet{bet10c} studied Kes 75 with their spectral evolution model.
There are two main differences between their model and ours.
First, while we model the magnetic field evolution assuming energy conservation (Equation (\ref{eq_magnetic-field})), they consider the adiabatic loss of the magnetic energy.
Second, they consider dynamical evolution of a PWN inside a SNR.
The age of the system is determined from the dynamical properties of system in their model, while we determine it from the spectral properties.
They assume that the age of the system is 650yr old, which is close to our result 700yr.
The parameters of the particle injection $p_1 = 1.7$ and $\gamma_{\rm b} = 8 \times 10^5$ are almost consistent with our model $p_1 = 1.6$ and $\gamma_{\rm b} = 2 \times 10^6$, but the high energy power-law index at injection $p_2 = 2.3$ is a little harder than our value $p_2 =2.5$.
The lower value of $p_2$ increases the calculated flux in the hard X-rays, but their fitted spectrum still underpredicts hard X-ray observation.
Their current magnetic field strength $\sim 30 \mu \rm G$ is almost consistent with our model $B_{\rm now} = 20 \mu \rm G$, but the adopted energy density of the ISRF $\sim 24 \rm eV/cm^3$ is very large.
However, because the assumed temperature of the ISRF $T_{\rm ISRF} = 1000 \rm K$ is high, Klein-Nishina effect significantly suppresses the IC/ISRF flux in their model. 

The uniqueness of Kes 75 PWN also appears in \citet{bet10c}.
Only for the case of Kes 75, the fraction of the magnetic energy injection $\dot{E}_{\rm mag}$ of the spin-down power $L(t)$ (corresponding to $\eta$ in our model) is almost two orders of magnitude lower than other young PWNe in both studies.
This relative smallness of $\eta$ compared with other young PWNe is likely to be real and is very interesting, although the absolute value of $\eta$ is different between \citet{bet10c} and the present work.
The difference of the absolute value of $\eta$ is most probably due to the difference of the magnetic field evolution model.

Lastly, we discuss the hard X-rays.
\citet{met08} argued that the hard X-ray emission detected by \textit{INTEGRAL} is dominated by the emission from the PWN, i.e., it is not from the pulsar.
However, it is difficult to reproduce the observed hard X-ray emission with the current magnetic field $B_{\rm now} = 20 \mu \rm G$.
If both the observed soft and hard X-ray spectra are fitted by power-law spectra as discussed in \citet{met08}, the synchrotron cooling break frequency $\nu_{\rm c}$ should be above $10^{19} \rm Hz$ ($B_{\rm now} \sim 5 \mu \rm G$ for $t_{\rm age} = 0.7 \rm kyr$) to reproduce the observed soft X-ray photon index $\Gamma_X < 2$.
This is difficult to realize for the same reason as for G21.5-0.9 because the observed $\gamma$-rays to X-rays flux ratio constrains the current magnetic field strength.
So that we think that the hard X-ray emission may have a pulsar origin.

\section{G0.9+0.1}\label{g0.9}
G0.9+0.1 is a composite supernova remnant and its PWN component is observed in radio \citep[][]{det08b},  X-rays \citep[][]{get01}, and TeV $\gamma$-rays \citep[][]{aet05}.
In X-rays, a jet-torus like structure is observed \citep[][]{get01, pet03}.
In radio, G0.9+0.1 consists of a compact PWN which has a radius of $\sim 1'$ and an outer SNR shell $\sim 4'$ in radius \citep[][]{det08b}.
No X-ray counterpart of the outer SNR shell is observed \citep[][]{pet03}.
The central pulsar of G0.9+0.1 (PSR J1747-2809) has recently been detected in radio, with a period $P = 5.22 \times 10^{-2} \rm sec$, its time derivative $\dot{P} = 1.56 \times 10^{-13} \rm sec \cdot sec^{-1}$ ($\tau_c = 5.3 \rm kyr$) and unknown braking index \citep[][]{cet09}.
So that we assume $n = 3$.
The actual age of G0.9+0.1 is unknown.
We consider two cases for the distance of G0.9+0.1: 8kpc (lower limit) and 13kpc (upper limit) given by \citet{cet09} as model 1 and 2, respectively.
The ISRF energy density of model 1 is larger than model 2, especially in optical band.
We assume that G0.9+0.1 PWN is a sphere of radius 2.3pc at $(r, z) \sim (0 {\rm kpc}, 15 {\rm pc})$ or 3.8pc at $(r, z) \sim (5 {\rm kpc}, 20 {\rm pc})$ for model 1 and 2, respectively.
The energy density of the ISRF is $(U_{\rm IR}, U_{\rm OPT}) = (1.6 {\rm eV/cm^3}, 15 {\rm eV/cm^3})$ for model 1 and $(U_{\rm IR}, U_{\rm OPT}) = (1.2 {\rm eV/cm^3}, 2.0 {\rm eV/cm^3})$ for model 2, respectively.

\subsection{Model 1}
\begin{figure}
\plotone{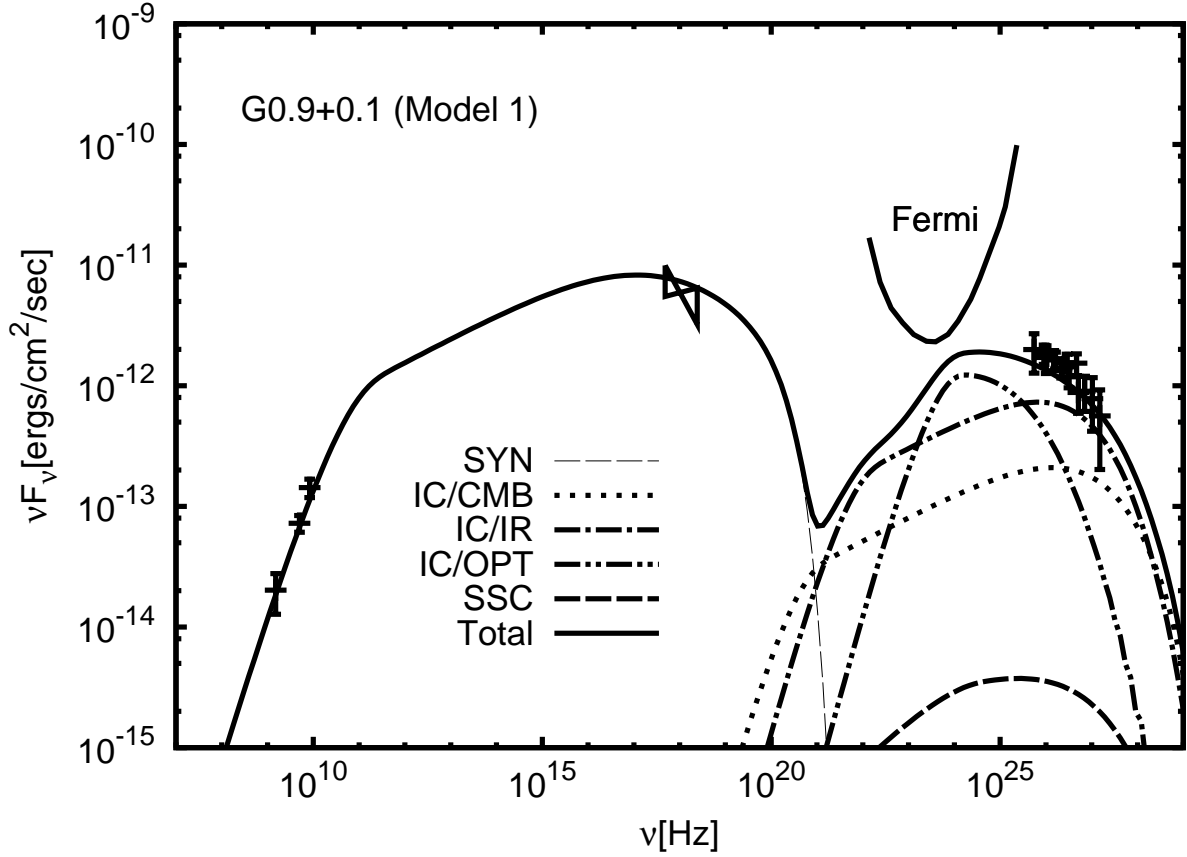}
\caption{Model spectrum of G0.9+0.1 at $t_{\rm age} = 2.0 \rm kyr$ for model 1, where the distance is taken to be 8kpc.
The observational data and the 1yr, 5$\sigma$ sensitivity for {\it Fermi} LAT are plotted.
However, this sensitivity curve may be more worse because G0.9+0.1 is significantly closer to the Galactic center than G21.5-0.9 on the sky.
The observed data are taken from \citet{det08b} (radio), \citet{get01} (X-ray), \citet{aet05} ($\gamma$-ray).
Used and obtained parameters are tabulated in Table \ref{tbl-1}.
\label{g0.9_model1_current}}
\end{figure}

Figure \ref{g0.9_model1_current} shows the model spectrum of G0.9+0.1 when the distance to G0.9+0.1 is taken to be 8kpc together with observational data.
We fit the data with the parameters $\eta = 3.0 \times 10^{-3}$, $t_{\rm age} = 2.0 \rm kyr$, $\gamma_{\rm max} = 8.0 \times 10^8$, $\gamma_{\rm b} = \gamma_{\rm min} =4.0 \times 10^4$, and $p_2 = 2.6$.
The observed radio spectrum is not fitted with the usual value of $p_1$ in a range of $p_1 > 1$ but we take $\gamma_{\rm min} = \gamma_{\rm b}$.
The fitted fraction parameter is close to that of the Crab Nebula, being just 0.6 times the value of the latter.
The fitted age $t_{\rm age} = 2.0 \rm kyr$ deviates from the characteristic age $\tau_c = 5.3 \rm kyr$, i.e., the current period is close to the initial period.
The expansion velocity $v_{\rm PWN} = 1120 \rm km / sec$ is comparable with that of G21.5-0.9 and model 2 of G54.1+0.3.
The pulsar parameters are $\tau_0 = 3.2 \rm kyr$, $L_0 \cdot \tau_0 = 1.2 \times 10^{49} \rm erg$ and the current total energy $E_{\rm tot}(2.0 \rm kyr) = 4.4 \times 10^{48} \rm erg$, which is almost a half of $L_0 \cdot \tau_0$.
The current magnetic field strength of G0.9+0.1 turns out to be $B_{\rm now} = 15 \mu \rm G$.

A single power-law injection well describes the observations.
The radio emission comes from the particles suffering from the adiabatic cooling and details will be discussed in Section \ref{discussion}.
The $\gamma$-ray emission is dominated by the IC/IR and IC/OPT.
The synchrotron cooling break frequency $\nu_{\rm c} \sim 5 \times 10^{16} \rm Hz$ corresponds to the flux peak.

\begin{figure}
\plottwo{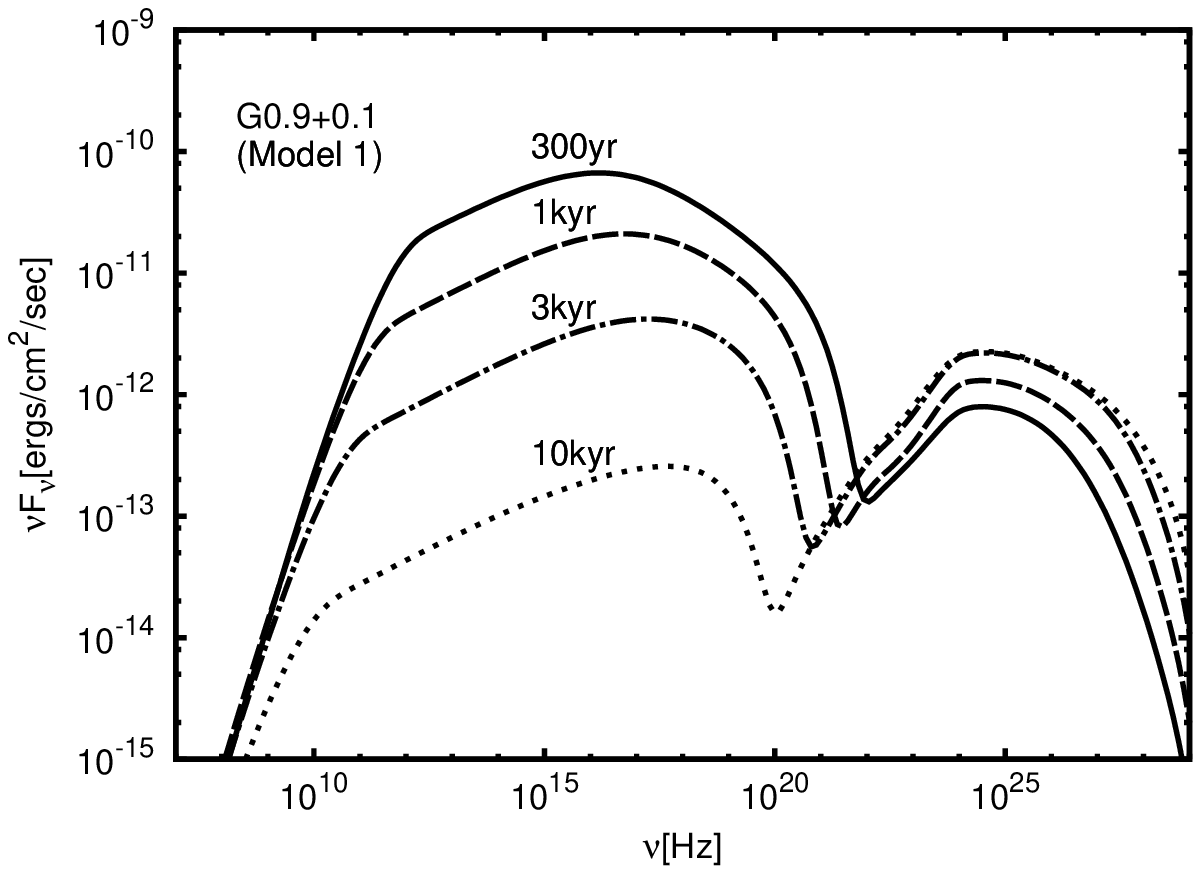}{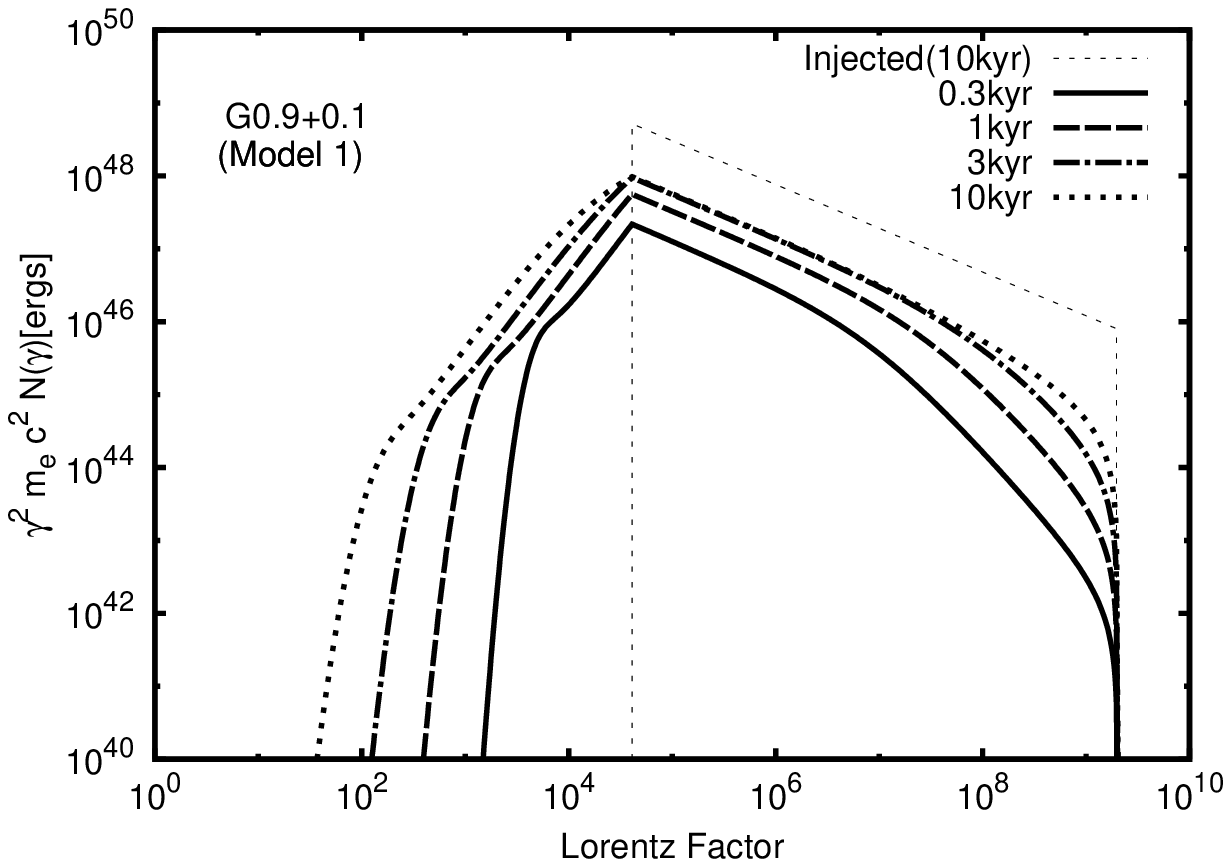}
\caption{Evolution of the emission spectrum (left panel) and the particle distribution (right panel) of G0.9+0.1 for model 1.
The total injected particles at an age of 10kyr (thin dotted line in right panel) shows that the injection spectrum is given by a single power-law distribution.
Note that the pileup feature is appeared in the right panel of the particle Lorentz factor $\gamma < 10^4$ for $t = 0.3 \rm kyr$.
This is made of numerical error and the particles in this feature do not contribute the emission spectrum of $\nu > 10^8 \rm Hz$.
\label{g0.9_model1_evolution}}
\end{figure}

Figure \ref{g0.9_model1_evolution} shows the evolution of the emission spectrum (left panel) and the particle distribution (right panel) of G0.9+0.1 with the use of the same parameters as in Figure \ref{g0.9_model1_evolution}.
While the synchrotron flux decreases, the inverse Compton flux increases with time.
This feature is the same as G21.5-0.9 because a spin-down time $\tau_0 = 3.2 \rm kyr$ is large.
In the right panel, we see how a single power-law distribution evolves with time, producing the low energy tail.

\subsection{Model 2}
\begin{figure}
\plotone{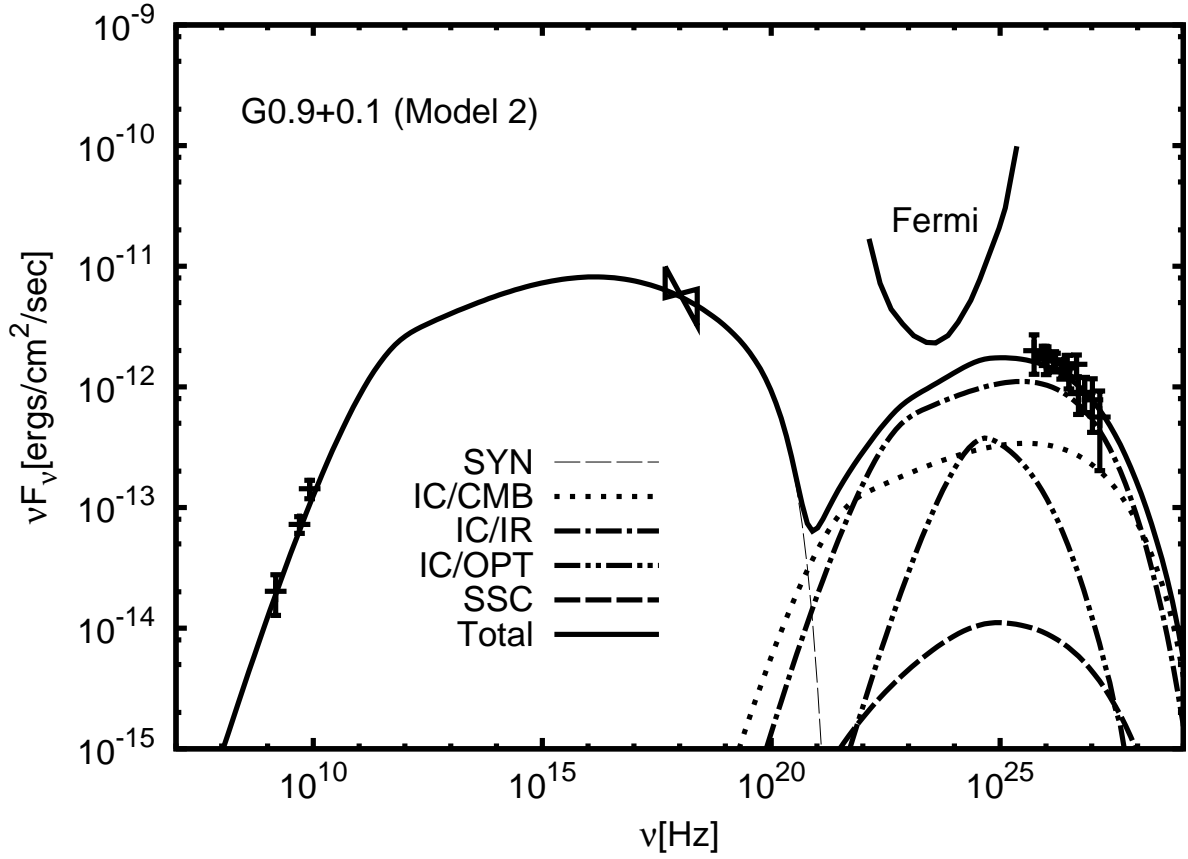}
\caption{Model spectrum of G0.9+0.1 at $t_{\rm age} = 4.5 \rm kyr$ for model 2, where the distance is taken to be 13kpc.
{\it Fermi} LAT sensitivity and the observed data are the same as in Figure \ref{g0.9_model1_current}.
\label{g0.9_model2_current}}
\end{figure}

Figure \ref{g0.9_model2_current} shows the model spectrum of G0.9+0.1 when the distance to G0.9+0.1 is taken to be 13kpc together with observational data.
We fit the data with the parameters $\eta = 1.0 \times 10^{-3}$, $t_{\rm age} = 4.5 \rm kyr$, $\gamma_{\rm max} = 1.0 \times 10^9$, $\gamma_{\rm b} = \gamma_{\rm min} = 1.0 \times 10^5$, and $p_2 = 2.6$.
The fraction parameter is similar to that of model 1.
The main difference from model 1 is in the fitted age of G0.9+0.1, which becomes more than twice that of model 1 because we need a larger current total energy $E_{\rm tot}$ is needed given the larger distance to the object.
The expansion velocity $v_{\rm PWN} = 830 \rm km / sec$ is a little smaller than model 1.
The pulsar parameters of G0.9+0.1 are $\tau_0 = 0.8 \rm kyr$ and $L_0 \cdot \tau_0 = 4.8 \times 10^{49} \rm erg$.
The current total energy $E_{\rm tot}(4.5 \rm kyr) = 4.2 \times 10^{49}$ is an order of magnitude larger than model 1.
Note that $t_{\rm age}$ is larger than $\tau_0$ for model 2 and vice versa for model 1.
The current magnetic field strength $B_{\rm now} = 12 \mu \rm G$ is similar to model 1.

The $\gamma$-ray emission is dominated by the IC/IR component.
Although the energy density of the ISRF is very different from model 1, the energy density of the magnetic field remains similar, because of the strong Klein-Nishina effect on the IC/OPT in model 1.
Note that the difference of the ISRF energy density between model 1 and 2 is mainly that in optical band.
The synchrotron cooling break frequency is $\nu_{\rm c} \sim 2 \times 10^{16} \rm Hz$.

\begin{figure}
\plottwo{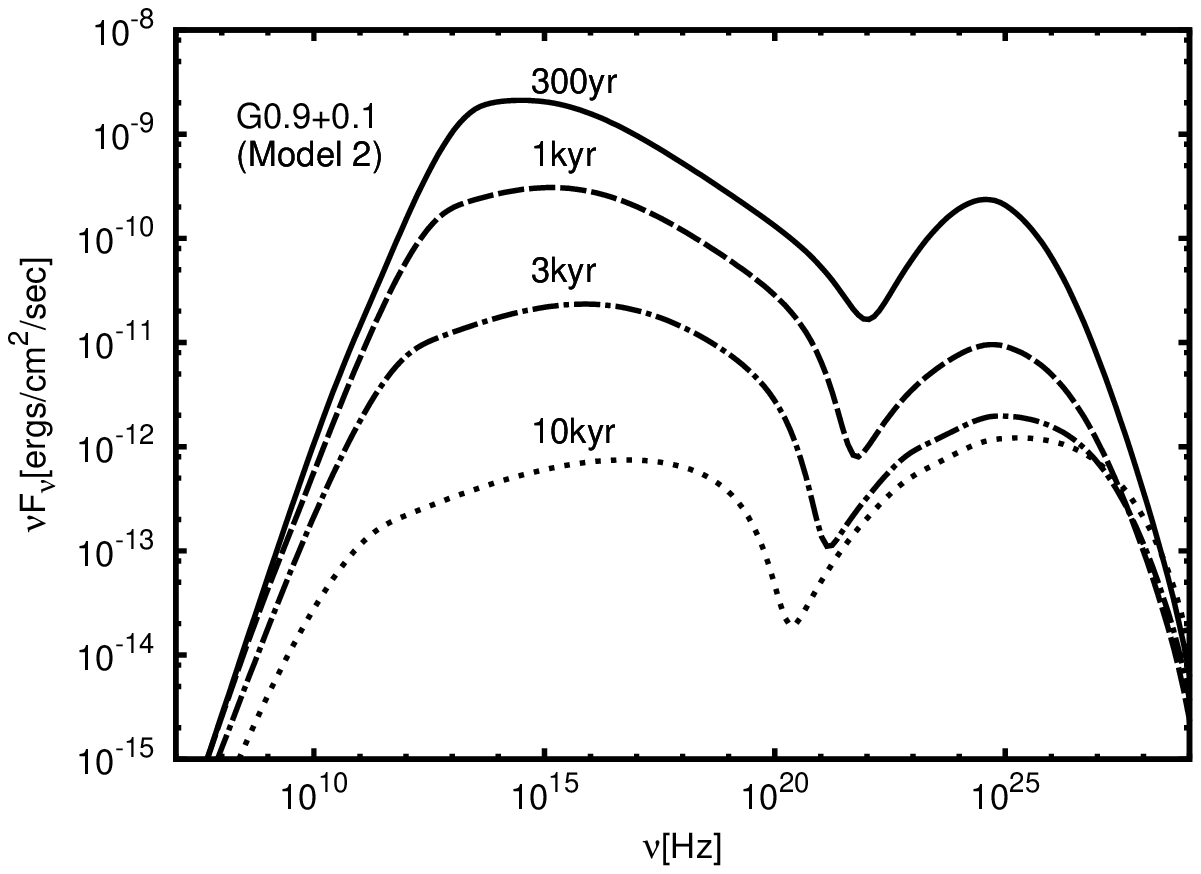}{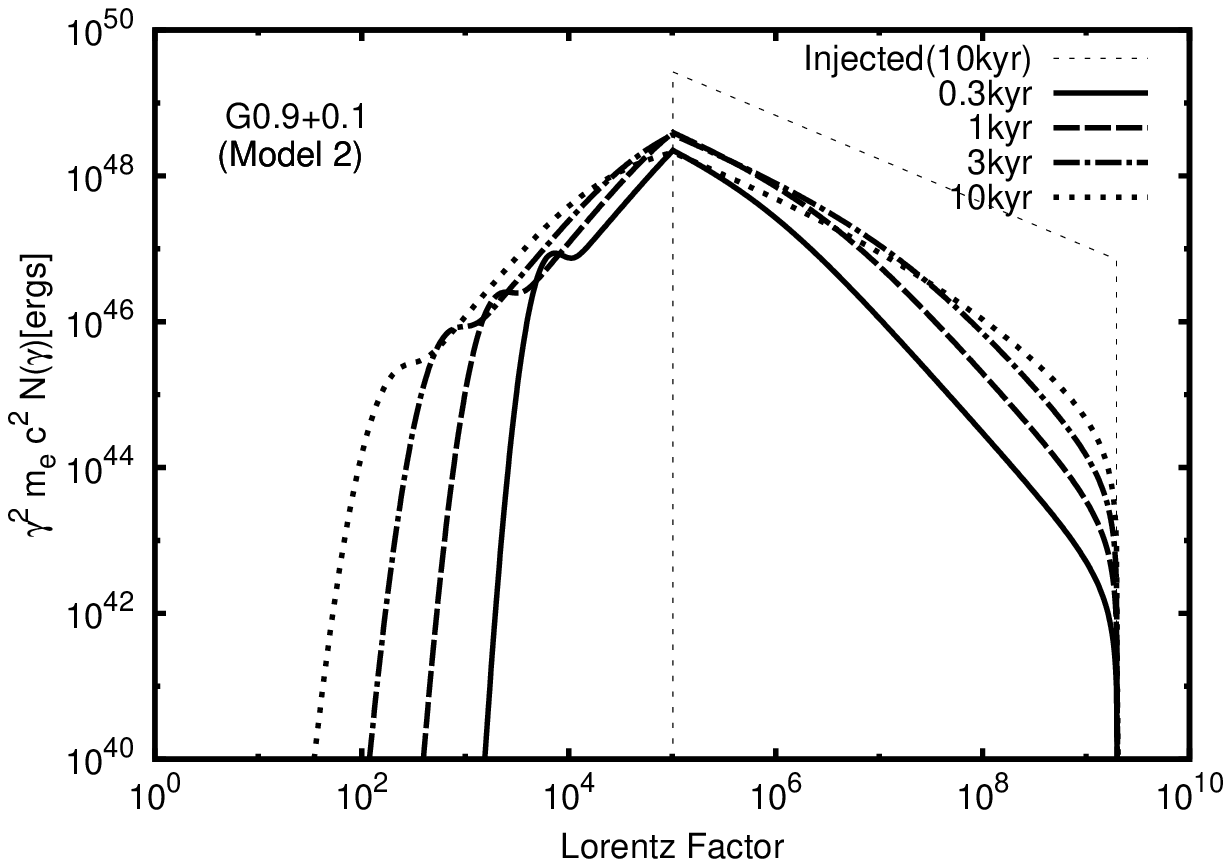}
\caption{Evolution of the emission spectrum (left panel) and the particle distribution (right panel) of G0.9+0.1 at 13kpc.
As seen from the right panel, the injection spectrum is a single power-law distribution.
Note that the pileup feature is also appeared in the right panel of the particle Lorentz factor $\gamma < 10^4$ for $t = 0.3 \rm kyr$ and is the same as the right panel of Figure \ref{g0.9_model1_evolution}.
\label{g0.9_model2_evolution}}
\end{figure}

Figure \ref{g0.9_model2_evolution} shows the evolution of the emission spectrum (left panel) and the particle distribution (right panel) of G0.9+0.1 with the use of the same parameters as in Figure \ref{g0.9_model2_evolution} (model 2).
Both the synchrotron flux and the IC flux decrease with time, and this behavior is the same as G54.1+0.3 shown in the left panel of Figure \ref{g54.1_model1_evolution}.
This different evolution in the left panels of Figure \ref{g0.9_model2_evolution} from in Figure \ref{g0.9_model1_evolution} arises from the difference between $t_{\rm age}$ and $\tau_0$ in the models.

\subsection{Discussion}
Both model 1 and 2 well reproduce the observed spectra and both models are acceptable.
The models differ in the IC/OPT flux, but the current observations cannot distinguish the models.
A low energy, power-law component for the injected particles, is not needed (see the injection spectrum of the right panels of Figures \ref{g0.9_model1_evolution} and \ref{g0.9_model2_evolution}), to reproduce the observed radio spectrum of G0.9+0.1.
This is because the adiabatic cooling of the injected particles can create the observed radio spectrum as will be discussed in Section \ref{discussion}.
However, we should note that no other PWNe which we have studied can be fitted by the single power-law injection of the particles.

\citet{fz10} studied the spectral evolution of G0.9+0.1.
They adopted the model in which G0.9+0.1 is located at the galactic center ($d = 8.5 \rm kpc$ in their model) with similar values of the ISRF energy densities $U_{\rm IR} = 0.5 \rm eV/cm^3$ and $U_{\rm opt} = 20 \rm eV/cm^3$ to our model 1.
The most interesting point in their model is that the particle distribution at injection is given by a relativistic Maxwellian plus single power-law distribution \citep[e.g.,][]{s08}, not a broken power-law distribution.
They could reproduce the observed spectrum of G0.9+0.1 with $p_2 = 2.5$ and $\gamma_{\rm b} \sim 4 \times 10^4$. 
The current magnetic field strength $8.1 \mu \rm G$ is about a half of our value $B_{\rm now} = 15 \mu \rm G$.
As will be discussed in Section \ref{discussion}, their conclusion is consistent with our study, but the existence of the relativistic Maxwellian component is not essential to reproduce the observed radio spectrum. 
We think that adiabatic cooling of the injected particles could also works in their model to reproduce the observed radio spectrum.

\section{DISCUSSIONS AND CONCLUSIONS}\label{dis_con}
We discuss first the dependence of emission spectrum on the adopted ISRF energy densities.
Next, we discuss about the difference and similarity of the fitted and derived parameters among five young TeV PWNe including the Crab Nebula.
We search for correlations between the central pulsar properties and the fitted parameters.

\subsection{Discussion}\label{discussion}
There is a possibility that the ISRF energy density is locally different from the mean values of the Galaxy as discussed in the case of model 2 of G54.1+0.3 and we see the $\gamma$-ray emission is dominated by the IC/ISRF except for the Crab Nebula.
To reproduce the observed power of the IC/ISRF, a larger ISRF energy density $U_{\rm ISRF}$ leads to a smaller current total energy of particles $(1 - \eta) E_{\rm tot}(t_{\rm age})$.
Accordingly, the fraction parameter $\eta$ needs to be larger to reproduce the observed power of the synchrotron radiation.
We showed it in our previous paper \citep[Section 3.3 of][]{tt10} that the power of the synchrotron radiation and that of the IC/ISRF roughly behave as $P_{\rm syn} \propto (1 - \eta) \eta E^2_{\rm tot}$ and $P_{\rm IC/ISRF} \propto (1 - \eta) E_{\rm tot} U_{\rm ISRF}$, respectively.
These relations lead to $E_{\rm tot} \propto U^{-1}_{\rm ISRF}$ and $\eta \propto U^2_{\rm ISRF}$ together with observed $P_{\rm syn}$ and $P_{\rm IC/ISRF}$.
As for $t_{\rm age}$, we find that a larger $U_{\rm ISRF}$ leads to a smaller $t_{\rm age}$ from the integration of Equation (\ref{eq_spin-down}).

However, Klein-Nishina effect makes the dependence of $E_{\rm tot}$ and $\eta$ on $U_{\rm ISRF}$ somewhat milder.
As seen from Table \ref{tbl-1}, the adopted $U_{\rm IR}$ in model 2 of G54.1+0.3 is four times larger than that in model 1, but $E_{\rm tot}$ of model 1 is about three times larger than that of model 2.
Accordingly, $\eta$ of model 2 is about seven times larger than that of model 1.
In section \ref{dis_kes75}, we considered how large $U_{\rm ISRF}$ is required in order for $\eta$ of Kes 75 to be as large as other PWNe.
We can estimate that more than an order of magnitude larger $U_{\rm ISRF}$ is required for model 1 and almost two order of magnitude larger $U_{\rm ISRF}$ is required for model 2.

For the fraction parameter $\eta$, all the young TeV PWNe have a value much smaller than unity and most of them are similar to each other.
We conclude that the fraction parameter $\eta$ in our model is typically a few $\times 10^{-3}$ for young TeV PWNe except for Kes 75.

For the break energy, $\gamma_{\rm b} \sim 10^{5 - 6}$ is found for all the young TeV PWNe.
Together with the minimum energy $\gamma_{\rm min}$ and the low energy power-law index at injection $p_1$, these parameters determine the supply rate of the particles from the pulsar wind and thus determine the pair multiplicity $\kappa$ inside the pulsar magnetosphere and the bulk Lorentz factor of the pulsar wind $\Gamma_{\rm w}$ \citep[][]{tt10, bet10c}.
For the typical values of the power-law indices at injection $1 < p_1 < 2 < p_2 < 3$, the particle number flux is estimated as $\dot{N}_{\rm inj}(t) \sim \gamma_{\rm min} Q_{\rm inj}(\gamma_{\rm min}, t)$ and the particle energy flux is estimated as $\dot{E}_{\rm part}(t) \sim \gamma^2_{\rm b} m_e c^2 Q_{\rm inj}(\gamma_{\rm b}, t)$.
We can estimate $\kappa$ as $\dot{N}_{\rm inj} = \kappa \dot{N}_{\rm GJ}$, where $\dot{N}_{\rm GJ}$ is the Goldreich-Julian number flux and $\Gamma_{\rm w}$ is estimated as $L(t) \sim \dot{E}_{\rm part}(t) \sim \dot{N}_{\rm inj}(t) \Gamma_{\rm w} m_e c^2$.
The fitted minimum energy ($\gamma_{\rm min} <$ a few $\times 10^3$)  gives a lower limit of $\kappa$ and an upper limit of $\Gamma_{\rm w}$.
However, the values of $\kappa$ and $\Gamma_{\rm w}$ of G0.9+0.1 is fixed because $\gamma_{\rm min} = \gamma_{\rm b}$.
Derived $\kappa$ and $\Gamma_{\rm w}$ for each PWN are listed in Table \ref{tbl-2}.
We find that the lower limit of $\kappa$ is larger than $10^4$ for all young TeV PWNe and that the upper limit of $\Gamma_{\rm w}$ is smaller than $\gamma_{\rm b}$ for the Crab Nebula and Kes 75 (model 1 and 2).
These quantities may be more constrained by the future lowest frequency radio observations, such as {\it LOFAR}, {\it ASKAP} and {\it SKA}.

The high energy power-law index at injection $p_2 \sim 2.5$ takes a very similar value for all five young TeV PWNe.
As stressed by \citet{bet10c}, spatially variable fluxes and photon indices of X-ray observations are not guaranteed to be reproducible in a one-zone broad band emission models.
For example, the hard X-ray observation of Kes 75 may suggest that the hard X-ray emission mainly comes from the central region of strong magnetic fields.
However, the result of $p_2 \sim 2.5$ for all the young TeV PWNe is still interesting.
It suggests that acceleration process at the pulsar wind termination shock and/or the cooling, advection and diffusion of the accelerated particles are common to young PWNe \citep[e.g.,][]{bet10d}.

The low energy power-law index at injection $p_1$, on the other hand, is different for each PWN and varies $1.0 \le p_1 \le 1.6$.
Moreover, in the case of G0.9+0.1, we either do not need a low energy component or it should be very hard ($p_1 \ll 1.0$).
On the assumption of an uniform PWN, this behavior of G0.9+0.1 is explained in the following way.
Low energy particles are mainly cooled by adiabatic cooling.
When we take into account only the adiabatic cooling, Green's function of the continuity equation of particles in energy space
\begin{equation}\label{eq_adiabatic}
\frac{\partial}{\partial t} G(\gamma, t) - \alpha \frac{\partial}{\partial \gamma} \left( \frac{\gamma}{t} G(\gamma, t) \right) = \delta(\gamma - \gamma_0) \delta(t - t_0) 
\end{equation}
becomes
\begin{equation}\label{eq_greens-function}
G(\gamma, t) = \frac{1}{\alpha} \cdot \gamma^{-1} \Theta(\gamma_0 - \gamma) \delta \left( \ln \frac{t}{t_0} + \frac{1}{\alpha} \ln \frac{\gamma}{\gamma_0} \right),
\end{equation}
where $\alpha$ takes into account an accelerated ($\alpha > 1$) or decelerated ($\alpha < 1$) expansion of the PWN ($R_{\rm PWN} \propto t^{\alpha}$) and $\Theta(x)$ is the Heaviside step function.
For simplicity, we consider the solution of the time-dependent injection $q(\gamma, t) = q_0 (t / t_1)^{- \beta} \delta(\gamma - \gamma_1) \Theta(t - t_1)$, where $q_0 = \rm{const.}$ is a normalization factor, $t_1$ is the time when the injection starts, $\gamma_1$ is the particle energy at injection and $\beta$ is the time dependence of injection.
The particle distribution $N(\gamma, t)$ is given by
\begin{equation}\label{eq_adiabatic-solution}
N(\gamma, t) = \frac{q_0 t}{\alpha \gamma_1} \cdot \left( \frac{\gamma}{\gamma_1} \right)^{\frac{1 - \beta}{\alpha} - 1} \left( \frac{t}{t_1} \right)^{- \beta} \Theta(\gamma_1 - \gamma) \Theta \left( \gamma - \left( \frac{t_1}{t} \right)^{\alpha} \gamma_1 \right).
\end{equation}
We consider the PWN expanding at a constant velocity $\alpha = 1$ and the particles whose energy is lower than $\gamma_1 = \gamma_{\rm b} = \gamma_{\rm min}$ for G0.9+0.1.
Equation (\ref{eq_adiabatic-solution}) gives the particle distribution $N \propto \gamma^{- \beta}$ for $\gamma < \gamma_1 = \gamma_{\rm min}$.
From Equation (\ref{eq_spin-down}), for $t \ll \tau_0$, $\beta$ is 0 while $\beta$ is $(n + 1) / (n - 1)$ for $t \gg \tau_0$.
However, for $t_{\rm age} \sim \tau_0$ $\sim O(\rm kyr)$ as in the both models of G0.9+0.1, $\beta$ smoothly varies from 0 to 2 for $n = 3$ on the time scale of $\tau_0$.
As seen in the right panels of Figures \ref{g0.9_model1_evolution} and \ref{g0.9_model2_evolution} (see thick dotted curves at 10 kyr), the particle distribution for $\gamma < \gamma_{\rm b}$ continuously changes from $0$ to $-2$ on the time scale of $\tau_0$.
Thus, the observed radio spectral index $\alpha_r = -0.18$ \citep[][]{det08b}, corresponding to the power-law index $\sim -1.4$ of particle distribution, can be almost reproduced by the adiabatic cooling.
Note that the slope of the particle distribution for $\gamma < \gamma_{\rm b}$ is slightly different between the right panels of Figures \ref{g0.9_model1_evolution} and \ref{g0.9_model2_evolution}, because $\tau_0$ is different.
Note also that the observed radio spectral index $\alpha_r = -0.18$ is clearly smaller than $\alpha_r = 1/3$ which corresponds to the low frequency tail of synchrotron radiation.

\begin{figure}
\plotone{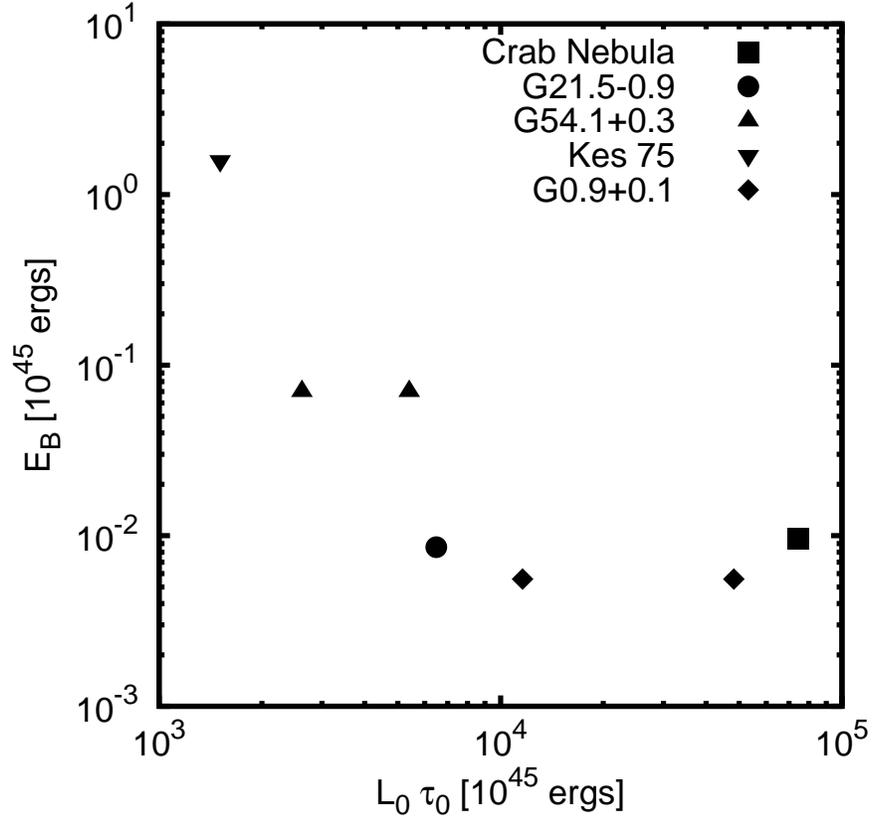}
\caption{The correlation between the initial rotational energy $L_0 \cdot \tau_0$ versus the magnetic energy $B^2_{\rm ini} R^3_{\rm pulsar} / 6$ of the central pulsars.
Model 1 of G54.1+0.3 and model 2 of Kes 75 are not plotted.
\label{rot-mag-energy}}
\end{figure}

Concerning the properties of the central pulsars, when the braking index $n$ is given, the spin-down evolution of the pulsar is characterized by two parameters, the initial spin-down power $L_0$ and the spin-down time $\tau_0$ in Equation (\ref{eq_spin-down}). 
In other words, the individualities of each pulsar come from these two quantities which theoretically represent the two parameters equivalent to the initial rotational energy and the magnetic energy of the pulsar.
In Figure \ref{rot-mag-energy}, we plot the correlation between the initial rotational energy $L_0 \cdot \tau_0$ versus the magnetic energy of the pulsar $E_{\rm B} = B^2_{\ast}R^3_{\ast} / 6$, where $B_{\ast} \propto \dot{P}^{1/2} P^{1/2}$ (assuming magnetic dipole radiation) and $R_{\ast} = 10^6 \rm cm$ are the surface dipole magnetic field and the radius of the pulsar, respectively.
Figure \ref{rot-mag-energy} shows an anticorrelation between $L_0 \cdot \tau_0$ and $E_{\rm B}$, although the statistics is rather poor and Kes 75 dominates in this anticorrelation.
We should mention that this anticorrelation remains unchanged even if we use the canonical value $n = 3$ for Kes 75, instead of we have used $n = 2.65$.
Although the canonical value $n = 3$ keeps $E_{\rm B}$ unchanged from its birth, when $n$ is not 3, it seems better to use the initial value of $E_{\rm B}$ ($E_{\rm B_0}$) rather than that at present time $E_{\rm B}(t_{\rm age})$.
However, for Kes 75 ($n = 2.65$), $E_{\rm B}(t_{\rm age})$ is only a factor of 1.3 smaller than $E_{\rm B_0}$.
The value of $n$ also changes $L_0 \cdot \tau_0$.
However, $L_0 \cdot \tau_0$ for $n = 2.65$ is only a factor of 1.4 larger than that for $n = 3$ with fixed $E_{\rm tot}(t_{\rm age})$, which we can almost independently determine from the observed power of the IC/ISRF.

\begin{figure}
\plottwo{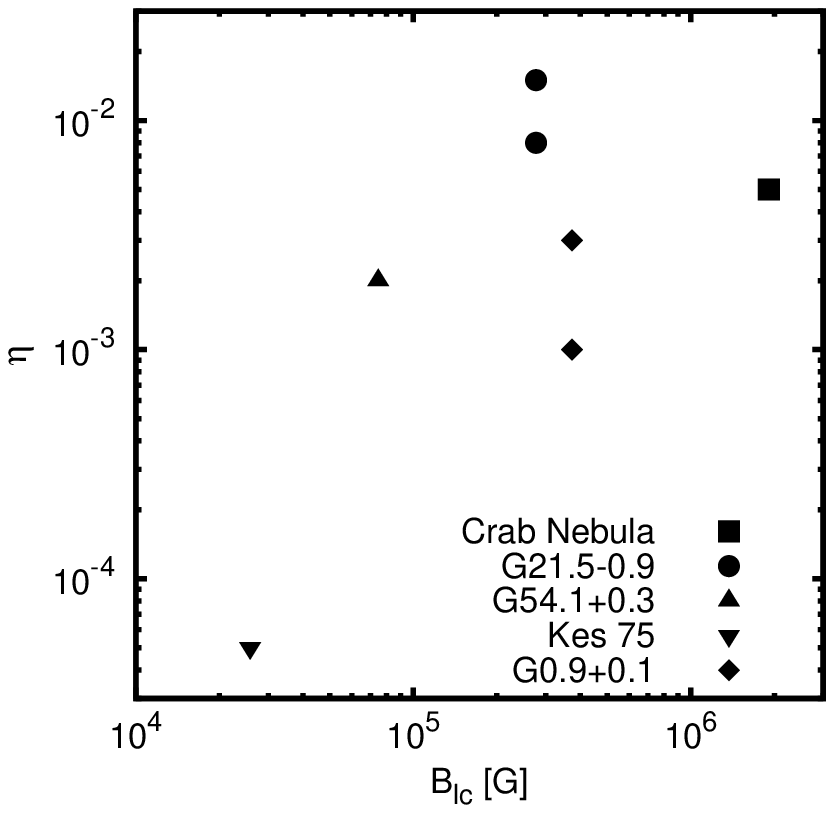}{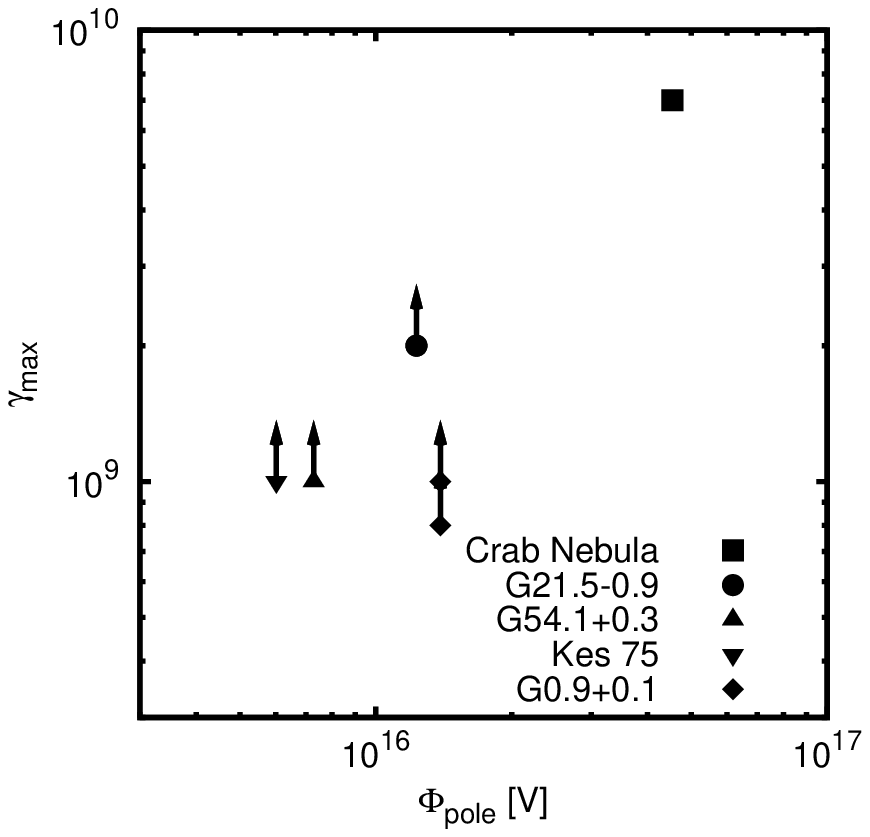}
\caption{The correlations between the fraction parameter $\eta$ versus the magnetic field strength at the light cylinder $B_{\rm lc}$ (left panel) and the maximum energy $\gamma_{\rm max}$ (lower limit except for the Crab Nebula) versus the potential difference at polar cap $\Phi_{\rm pole}$ (right panel).
Model 1 of G54.1+0.3 and model 2 of Kes 75 are not plotted.
\label{eta-max}}
\end{figure}

We also search for correlations between the parameters of the injection spectrum and the pulsar parameters.
The pulsar parameters include the spin-down power $L(t) \propto \dot{P} / P^3$, the surface dipole magnetic field $B_{\ast} \propto \dot{P}^{1/2} P^{1/2}$, the potential difference at the polar cap $\Phi_{\rm pole} \propto \dot{P}^{1/2} / P^{3/2}$ and the light cylinder magnetic field $B_{\rm lc} \propto B_{\ast} (R_{\ast} \Omega / c)^3 \propto \dot{P}^{1/2} / P^{5/2}$, where $\Omega$ is the current angular velocity of the pulsar.
In the left panel of Figure \ref{eta-max}, we plot the correlation between the fraction parameter $\eta$ versus the light cylinder magnetic field $B_{\rm lc}$.
Although it seems to show some correlation, when we ignore Kes 75, the correlation is insignificant and the values of $\eta$ for the other objects spread only in a range $10^{-3}$ -- $10^{-2}$.
In the right panel of Figure \ref{eta-max}, we plot the correlation between the maximum energy $\gamma_{\rm max}$ versus the potential difference at the polar cap $\Phi_{\rm pole}$.
Because $e \Phi_{\rm pole}$ gives the maximum available electric energy of the pulsar to accelerate particle, this correlation is expected as mentioned by \citet{bet10c}.
However, we do not find a significant correlation.
This may be partly because only an upper limit of $\gamma_{\rm max}$ is obtained, except for the Crab Nebula.
For other combinations of the parameters, we do not find any significant correlations and we do not show them here.

Lastly, we discuss about the age $t_{\rm age}$ and the expansion velocity $v_{\rm PWN}$ of PWNe.
Our spectral evolution model can estimate the age of the central pulsar in a fairly reliable way from the observed power of the IC/ISRF.
In contrast, the characteristic age $\tau_{\rm c}$ of a pulsar is not guaranteed to match the age of young pulsars, when the spin-down time $\tau_0$ is close to the age of the pulsar $t_{\rm age}$ as seen in Equation (\ref{eq_age_tc_t0}).
For example, $\tau_{\rm c}$ of the central pulsars inside the Crab Nebula and G21.5-0.9 are 1.2kyr and 4.8kyr, but $t_{\rm age}$ fitted with our model and the ages estimated from the observed expansion rate of the synchrotron nebula are both $\sim$ 1.0kyr.

We discuss about the properties of the SN explosion which creates the pulsars with obtained $v_{\rm PWN}$ and $E_{\rm tot}$. 
Simply, we expect the relation $v_{\rm PWN} \sim (E_{\rm tot}(t) / E_{\rm SN} )^{0.2} V_{\rm SN}$, where $E_{\rm SN}$ is the energy of SN ejecta and $V_{\rm SN}$ is the velocity of the front of the freely expanding ejecta.
We derive this relation from the radius of PWNe $R_{\rm PWN}$ in a freely expanding SN ejecta estimated as $R_{\rm PWN}(t) \sim ( E_{\rm tot}(t) / E_{\rm SN} )^{0.2} V_{\rm SN} \it{t}$ by \citet{vet01}. 
We do not find the relation $v_{\rm PWN} \propto E^{0.2}_{\rm tot}$, which suggests that the values of $E_{\rm SN}$ and $V_{\rm SN}$ are not common to the PWNe. 
On the other hand, because the derived values of $v_{\rm PWN}$ and $E^{0.2}_{\rm tot}$ differ by a factor of a few, the combination of $V_{\rm SN} / E^{0.2}_{\rm SN}$ also ranges within a factor of a few.

\subsection{Conclusions}\label{conclusion}
In this paper, we apply our spectral evolution model to four young TeV PWNe, G21.5-0.9, G54.1+0.3, Kes 75, and G0.9+0.1.
We have succeeded in reproducing many observed properties of these PWNe based on this rather simplified one-zone model.

The current observed spectra of all four young TeV PWNe are reconstructed with small values of the fraction parameter $\eta \ll 1$ as well as the Crab Nebula, i.e., the magnetic energy of these PWNe accounts for a very small fraction of the current total energy injected into the PWN $E_{\rm tot}(t_{\rm age})$.
The fitted fraction parameters are typically a few $\times 10^{-3}$ except for Kes 75.
The fraction parameter of the peculiar object Kes 75 is more than two orders of magnitude smaller than the typical value.

The TeV $\gamma$-ray emission from the young TeV PWNe is dominated by IC/ISRF.
Since the energy density of the local ISRF around the objects is somewhat uncertain, it is important to take into account its effect as considered in model 2 of G54.1+0.3.
On the other hand, the $\gamma$-ray emission at early phase of their evolution (e.g., $t_{\rm age} <$ 300yr) is always SSC dominant because the magnetic energy density of the PWN is much larger than the local ISRF energy density.

A broken power-law injection of particles well reproduces the observed spectrum from radio through TeV $\gamma$-ray, except the case of G0.9+0.1 where we do not need the low energy component.
The observed spectrum of G0.9+0.1 in radio is created by the adiabatic cooling of the high energy component of the injected particles.

The fitted break energy $\gamma_{\rm b} \sim 10^{5 - 6}$ is rather common.
On the other hand, the fitted maximum energy $\gamma_{\rm max}$ is a lower limit (more than $\sim 10^9$) and the fitted minimum energy $\gamma_{\rm min}$ is an upper limit (less than $\sim 10^4$).
The high energy power-law index at injection $p_2 \sim 2.5$ is common for all young TeV PWNe, while the low energy power-law index at injection $p_1$ varies in the range $1.0 \le p_1 \le 1.6$.

The fitted parameters of the injection spectrum in Equation (\ref{eq_injection}) gives a lower limit of the pair multiplicity $\kappa$, which turns out to be more than $10^4$.
We have also estimated an upper limit of the bulk Lorentz factor of the pulsar wind $\Gamma_{\rm w}$.
For G21.5-0.9 and G54.1+0.3, the upper limit of $\Gamma_{\rm w}$ is still consistent with $\gamma_{\rm b}$, but the obtained $\Gamma_{\rm w}$ of Kes 75 is clearly less than $\gamma_{\rm b}$.
The latter feature is the same as for the Crab Nebula in our previous work $\Gamma_{\rm w} < 10^4 < \gamma_{\rm b} \sim 10^6$.
On the other hand, for G0.9+0.1, the value of $\Gamma_{\rm w}$ is similar to $\gamma_{\rm b}$ without uncertainty in our model.

From the fitted age, we can derive the initial spin-down luminosity and the spin-down time of the central pulsar.
We take the initial rotational energy $L_0 \cdot \tau_0$ and the magnetic energy $E_{\rm B}$ of the pulsar as two independent parameters which characterize the spin-down evolution of the pulsar and search for a correlation between them.
They seem to be anticorrelated, although the statistics is rather poor.
We also search for correlations between the fraction parameter and the parameters of the injection spectrum versus the central pulsar properties.
However, we find no significant correlations.

\acknowledgments
We are grateful to Y. Ohira for useful discussions.
We thank the anonymous referee for his/her meticulous reading of the manuscript and helpful comments. 
This work is partly supported by KAKENHI (F. T. , 20540231)

\clearpage

\clearpage

\begin{deluxetable}{cccccccccc}
\tabletypesize{\tiny}
\rotate
\tablecaption{Adopted parameters and fitted and derived parameters to reproduce the current observed spectrum.
\label{tbl-1}}
\tablewidth{0pt}
\tablehead{
\colhead{Symbol}     & \colhead{Crab\tablenotemark{a}}      & \colhead{G21.5-0.9} & \colhead{G21.5-0.9\tablenotemark{b}}                               & 
\colhead{G54.1+0.3}  & \colhead{G54.1+0.3\tablenotemark{c}} & \colhead{Kes 75}    & \colhead{Kes 75\tablenotemark{d}}                                  & 
\colhead{G0.9+0.1}   & \colhead{G0.9+0.1\tablenotemark{d}}  \\
Model                &                            & 1         & 2                 & 1                          & 2         & 1                 & 2                          & 1        & 2                 \\
\hline
\multicolumn{10}{c}{Adopted Parameters}
}

\startdata

$d$(kpc)                                        & 2.0      & 4.8      & 4.8      & 6.2     & 6.2      & 6.0     & 10.6     & 8.0     & 13      \\
$R_{\rm PWN, now}$(pc)                          & 1.8      & 1.0      & 1.0      & 1.8     & 1.8      & 0.29    & 0.50     & 2.3     & 3.8     \\
$\textit{P}$(msec)                              & 33.1     & 61.9     & 61.9     & 136     & 136      & 326     & 326      & 52.2    & 52.2    \\
$\dot{\textit{P}}(10^{-13})$ & 4.21     & 2.02     & 2.02     & 7.51    & 7.51     & 70.8    & 70.8     & 1.56    & 1.56    \\
$\textit{n}$                                    & 2.51     & 3.0      & 3.0      & 3.0     & 3.0      & 2.65    & 2.65     & 3.0     & 3.0     \\
$U_{\rm IR}$($\rm eV / cm^3$)                   & ---      & 1.0      & 1.0      & 0.5     & 2.0      & 1.2     & 1.0      & 1.6     & 1.2     \\
$U_{\rm OPT}$($\rm eV / cm^3$)                  & ---      & 2.0      & 2.0      & 0.5     & 0.5      & 2.0     & 2.0      & 15      & 2.0     \\

\cutinhead{Fitted Parameters}

$\eta$($10^{-3}$)                               & 5.0      & 15       & 8.0      & 0.3     & 2.0      & 0.05    & 0.006    & 3.0     & 1.0     \\
$\textit{t}_{\rm age}$(kyr)                     & 0.95     & 1.0      & 1.0      & 2.3     & 1.7      & 0.7     & 0.88     & 2.0     & 4.5     \\
$\gamma_{\rm max}$($10^9$)                      & 7.0      & $>$ 2.0  & $>$ 2.0  & $>$ 1.0 & $>$ 1.0  & $>$ 1.0 & $>$ 0.8  & $>$ 0.8 & $>$ 1.0 \\
$\gamma_{\rm b}$($10^5$)                        & 6.0      & 1.2      & 0.7      & 3.0     & 1.8      & 20      & 50       & 0.4     & 1.0     \\
$\gamma_{\rm min}$($10^3$)                      & $<$ 0.1  & $<$ 3.0  & $<$ 3.0  & $<$ 20  & $<$ 20   & $<$ 5.0 & $<$ 5.0  & ---     & ---     \\
$\textit{p}_1$                                  & 1.5      & 1.0      & 1.0      & 1.2     & 1.2      & 1.6     & 1.4      & ---     & ---     \\
$\textit{p}_2$                                  & 2.5      & 2.55     & 2.5      & 2.55    & 2.55     & 2.5     & 2.5      & 2.6     & 2.6     \\

\cutinhead{Derived Parameters}

$\textit{v}^{}_{\rm PWN}$(km/sec)                 & 1800     & 980      & 980      & 770     & 1040     & 420     & 560      & 1120    & 830     \\
$B_{\rm now}$($\mu \rm G$)                      & 85       & 64       & 47       & 6.7     & 10       & 20      & 24       & 15      & 12      \\
$\tau_0$(kyr)                                   & 0.7      & 3.9      & 3.9      & 0.6     & 1.2      & 0.2     & 0.003    & 3.2     & 0.8     \\
$L_0 \cdot \tau_0$($10^{48} \rm erg$)          & 74       & 6.5      & 6.5      & 5.4     & 2.6      & 1.5     & 210      & 12      & 48      \\

\enddata

\tablenotetext{a}{Results are taken from \citet{tt10}.}
\tablenotetext{b}{All the adopted parameters are the same as model 1 of G21.5-0.9, but ignoring the observation in infrared.}
\tablenotetext{c}{Assumed $U_{\rm IR}$ is different with model 1 of G54.1+0.3.}
\tablenotetext{d}{Assumed distances to the objects are different with model 1 of Kes 75 and G0.9+0.1, respectively.}
\end{deluxetable}

\begin{deluxetable}{cccccccccc}
\tabletypesize{\scriptsize}
\rotate
\tablecaption{The derived pair multiplicity and bulk Lorentz factor.
\label{tbl-2}}
\tablewidth{0pt}
\tablehead{
\colhead{Symbol}    & \colhead{Crab}      & \colhead{G21.5-0.9} & \colhead{G21.5-0.9} & \colhead{G54.1+0.3} &
\colhead{G54.1+0.3} & \colhead{Kes 75}    & \colhead{Kes 75}    & \colhead{G0.9+0.1}  & \colhead{G0.9+0.1} \\
Model                    &          &       1  &       2  &       1 &       2  &       1 &       2  &       1 &       2 
}
\startdata
$\kappa$($10^4$)         & $>$ 420  & $>$ 13   & $>$ 19   & $>$ 3.7 & $>$ 5.2  & $>$ 2.8 & $>$ 0.85 & 8.3     & 3.4     \\
$\Gamma_{\rm w}$($10^5$) & $<$ 0.07 & $<$ 0.67 & $<$ 0.53 & $<$ 2.1 & $<$ 0.91 & $<$ 1.4 & $<$ 4.6  & 1.1     & 2.7     \\

\enddata

\end{deluxetable}

\end{document}